\documentclass[aip,jcp,reprint,superscriptaddress, citeautoscript]{revtex4-1}

\pdfoutput=1

\usepackage{amsmath,amsfonts,amssymb,bm,graphicx}
\usepackage{xcolor}

\makeatletter
\def\@bibdataout@aip{
 \immediate\write\@bibdataout{
 @CONTROL{
   aip41Control, author="08",editor="1",pages="0",title="0",year="1"
 }}
 \if@filesw
  \immediate\write\@auxout{\string\citation{aip41Control}}
 \fi
}
\makeatother

\newcommand{\updelta}{\delta}

\usepackage{cmap} % to allow searching through ligatures
\usepackage{microtype} % microtypography, for visually more pleasant columns

\usepackage{mleftright} % this allows us to use the correct spacing between an operator and a bracketed group appearing immediately after it (so we replace e.g. \exp\left(\right) with \exp\mleft(\mright) to give consistent horizontal spacing)

\usepackage{helvet}

\usepackage{flafter}

\newcommand{\boldnabla}{\nabla}

\usepackage{marginnote}

\begin{document}

\newcommand{\LL}{\mathbb{L}}
\newcommand{\HH}{\mathbb{H}}
\newcommand{\TT}{\mathcal{T}}
\newcommand{\ee}{\mathrm{e}}

\newcommand{\jj}{A}

\newcommand{\ddiv}{\operatorname{div}}

\newcommand{\rhobar}{\overline{\rho}}

\newcommand{\beq}{\begin{equation}}
\newcommand{\eeq}{\end{equation}}

\newcommand{\rlj}[1]{{\color{blue}#1}}

\newcommand{\Lb}{{\bm{L}}}
\newcommand{\pb}{{\bm{p}}}
\newcommand{\eb}{{\bm{e}}}
\newcommand{\rb}{{\bm{r}}}
\newcommand{\Rb}{{\bm{R}}}
\newcommand{\nb}{{\bm{n}}}
\newcommand{\fb}{{\bm{f}}}
\newcommand{\Fb}{{\bm{F}}}

\newcommand{\eps}{\varepsilon}

\newcommand{\Qpar}{\mathcal{Q}_\parallel}
\newcommand{\EEL}{\mathcal{E}_L}
\newcommand{\EEP}{\mathcal{E}_P}

\newcommand{\bfb}{\overline{\bm{f}}}

\newcommand{\Ac}{\mathcal{A}}
\newcommand{\PP}{\mathcal{P}}
\newcommand{\KK}{\mathcal{K}}
\newcommand{\KKb}{\overline{\mathcal{K}}}

\newcommand{\alphat}{\tilde\alpha}   

\newcommand{\kappat}{\kappa}
\newcommand{\lambdat}{\lambda}

\newcommand{\eqnRef}[1]{Eq.~\eqref{#1}}

\newcommand{\figrefsub}[2]{\ref{#1}#2}

\raggedbottom

\title{Microscopic analysis of thermo-orientation in systems of off-centre Lennard-Jones particles}

\begin{abstract} 
When fluids of anisotropic molecules are placed in temperature gradients, the molecules may align themselves along the gradient: this is called thermo-orientation.
We discuss the theory of this effect in a fluid of particles that interact by a spherically symmetric potential, where the particles' centres of mass do not coincide with their interaction centres.
Starting from the equations of motion of the molecules, we show how a simple assumption of local equipartition of energy can be used to predict the thermo-orientation effect, recovering the result of Wirnsberger \textit{et al.} [\textit{Phys.\ Rev.\ Lett.}~\textbf{120}, 226001 (2018)].
Within this approach, we show that for particles with a single interaction centre, the thermal centre of the molecule must coincide with the interaction centre.
The theory also explains the coupling between orientation and kinetic energy that is associated with this non-Boltzmann distribution.
We discuss deviations from this local equipartition assumption, showing that these can occur in linear response to a temperature gradient.
We also present numerical simulations showing significant deviations from the local equipartition predictions, which increase as the centre of mass of the molecule is displaced further from its interaction centre.
\end{abstract}

\author{Robert L. Jack}
\affiliation{Department of Chemistry, University of Cambridge, Lensfield Road, Cambridge CB2 1EW, United Kingdom}
\affiliation{Department of Applied Mathematics and Theoretical Physics, University of Cambridge, Wilberforce Road, Cambridge CB3 0WA, United Kingdom}
\author{Peter Wirnsberger}
\thanks{current employer: DeepMind, London, United Kingdom.}
\affiliation{Department of Chemistry, University of Cambridge, Lensfield Road, Cambridge CB2 1EW, United Kingdom}
\author{Aleks Reinhardt}
\affiliation{Department of Chemistry, University of Cambridge, Lensfield Road, Cambridge CB2 1EW, United Kingdom}

\maketitle

\section{Introduction}

The responses of fluids to non-equilibrium forcing can be rich and surprising.
For example, temperature gradients can drive thermophoretic motion~\cite{Duhr2006, Burelbach2018, Ganti2017} as well as instabilities to convection~\cite{Rosenblat1971}.
For fluids whose molecules lack inversion symmetry, one may observe spontaneous alignment of molecules along a thermal gradient~\cite{Romer2012, Daub2016, Lee2016, Wirnsberger2018, Gardin2019,olarte2018}; this is known as the thermo-orientation effect.
If the molecules in the fluid are polar, such a spontaneous orientation results in an emergence of electrical polarisation, or `thermopolarisation', as found in computer simulations of water~\cite{Bresme2008, Muscatello2011a, Armstrong2013, Iriarte-Carretero2016, Wirnsberger2016} and other fluids~\cite{Daub2014, Lee2016, Wirnsberger2017, Wirnsberger2018}.
Thermopolarisation and thermo-orientation effects are thought to have a common origin~\cite{Romer2012, Lee2016, Wirnsberger2018}; in the following, we consider non-polar molecules without any electrostatic interactions, so our analysis is restricted to the thermo-orientation effect.
However, we anticipate the principles described here can also be applied to thermopolarisation.

The analysis of such non-equilibrium effects is challenging: in contrast to equilibrium situations, methods that start from the Gibbs--Boltzmann distribution can no longer be applied, and the standard methods of equilibrium statistical physics lose much of their power.
If deviations from equilibrium are small, then one may exploit ideas of non-equilibrium thermodynamics~\cite{Groot1984}, but this is a macroscopic theory, so its predictive power for molecular properties is limited.

Alternatively, one may use methods based on kinetic theory, starting from microscopic equations of motion and adopting a mechanical approach based on the balance of forces and torques.  This leads to the hierarchy of equations due to Bogolyubov, Born, Green, Kirkwood and Yvon (BBGKY)~\cite{hansen-book}.  These equations provide a detailed microscopic description of the non-equilibrium steady state, but their analysis is feasible only if the complexity of the many-particle system can be simplified in some way.
 In gases, one may attack these equations directly by estimating the effects of collisions on particle positions and momenta~\cite{chapman-book}.  In liquids, one more commonly works with equations of motion for hydrodynamic variables (as in the approach of Irving and Kirkwood~\cite{Irving1950}), which can be closed by means of constitutive equations.
%For example, analysis of equilibrium systems requires a closure of the hierarchy of equations due to Bogolyubov, Born, Green, Kirkwood and Yvon (BBGKY)~\cite{hansen-book}.
Recently, methods that derive fluid properties directly from equations of motion have seen a resurgence of interest in non-equilibrium systems, particularly in active matter~\cite{Takatori2014,Solon2015-natphys,Winkler2015,Speck2016,Steffenoni2017,Klymko2017}, including methods based explicitly on force and torque balance~\cite{Takatori2014,Steffenoni2017,Klymko2017}.
Here, we use these methods to analyse thermo-orientation.

A recent article~\cite{Wirnsberger2018} presented a theory which can predict the degree of thermo-orientation based on molecular parameters and the equation of state of a (non-polar) reference fluid.
The theory starts from a thermodynamic perspective and its derivation requires several assumptions about the response of individual molecules to `ideal' (thermodynamic) forces.
In this work, we analyse these phenomena via the equations of motion of individual particles.
Our approach recovers the results of Ref.~\onlinecite{Wirnsberger2018} and also accounts for some deviations between theory and simulation seen in that work.
Moreover, our results provide a deeper understanding of the nature of the non-equilibrium steady state, including formulae for a (non-Boltzmann) probability distribution of single-particle properties.

The form of the paper is as follows.
Sec.~\ref{sec:model} defines the model and states is governing equations.  Sec.~\ref{sec:theory} describes the theory and the main physical insights that it provides.
In Sec.~\ref{sec:numerics}, we present numerical results, while Sec.~\ref{sec:conc} contains our conclusions.
Technical aspects of the theoretical calculations are included in appendices.

\section{Model}
\label{sec:model}

\subsection{Equations of motion for particles}

We focus on the simplest model for thermo-orientation that was considered in Ref.~\onlinecite{Wirnsberger2018}.  
Consider a fluid of identical particles interacting by Lennard-Jones (LJ) interactions.
We work in $d=3$ dimensions, although our results also apply for $d=2$.
%The centre of mass of particle $i$ is at position $\rb_i$, which is displaced by a distance $\alpha$ from the LJ interaction centre.
%That is, 
Particle $i$ has an orientation that is encoded in a unit vector $\eb_i$, and its LJ interaction centre is displaced by $\alpha\eb_i$ from the centre of mass.  Hence,
%and 
the interaction potential between particles $i$ and $j$ is $V(|\bm{R}_i - \bm{R}_j|)$, with
$\Rb_i=\rb_i+\alpha\eb_i$
and
$V(r)=4\eps[(\sigma/r)^{12} - (\sigma/r)^6]$, where $\sigma$ is the particle diameter and $\eps$ is the interaction energy.
Each particle has mass $m$ and moment of inertia $I$, which we assume to be a scalar.
We also define a dimensionless `molecular shape parameter' $\chi = m\alpha^2/I$.
In our numerical simulations, the moment of inertia is $I=m\sigma^2/10$, corresponding to a spherical mass distribution with diameter $\sigma$.  %, and  $\alpha=-\sigma/4$, consistent with Ref.~\onlinecite{Wirnsberger2018}.
%This corresponds to a molecular shape parameter of $\chi=5/8$.
The temperature $T$ is measured in units such that Boltzmann's constant $k_{\rm B}=1$.

\begin{figure}
\includegraphics[width=\linewidth]{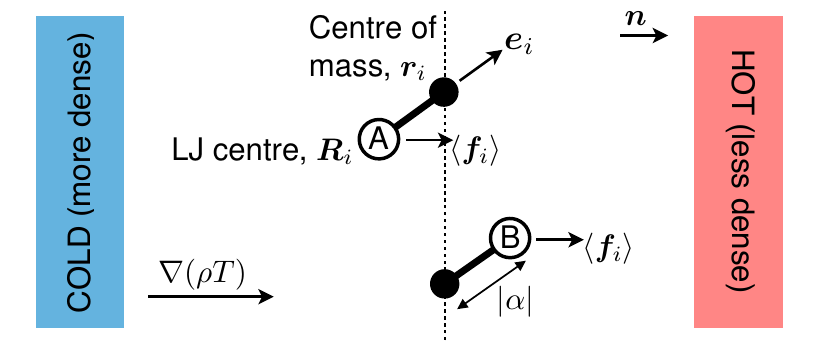}
\caption{Two molecules in a thermal gradient, without an imposed body force ($F^\text{ext}=0$).
Black circles indicate centres of mass and white circles indicate LJ centres.  The vector {from} the centre of mass {to} the LJ centre is $\alpha\eb_i$. 
We show the case $\alpha<0$, consistent with our numerical work.
The particles feel an average force to the right (as indicated), which acts at their LJ centre, and causes a torque that tends to increase the probability of configuration $B$.
However, the molecular orientation is coupled with its kinetic energy, which also affects the orientational statistics, as discussed in the main text.
The net result is that configuration $A$ is more probable than configuration $B$ in the system that we consider.  
(In cases where an external body force $F^\text{ext}$ is applied, it is assumed to act at the centre of mass.)
}
\label{fig:pic}
\end{figure}

In Fig.~\ref{fig:pic}, we show a schematic of the thermo-orientation effect.
The fluid is coupled to a hot and a cold reservoir.
After an initial transient period that depends on initial conditions, the system converges to a steady state in which there is a temperature gradient $\boldnabla T$.
Let $\nb$ be a unit vector parallel to $\boldnabla T$ and let
\beq
\Ac_T = \frac{1}{T} \left( \nb\cdot \boldnabla T\right)
 \label{equ:AcT-def}
\eeq
be the normalised (scalar) temperature gradient.
In our simulations, $\nb$ is the unit vector along the $z$ axis.
Following Ref.~\onlinecite{Wirnsberger2018}, our theory also includes an external body force of magnitude $F^\text{ext}$ parallel to $\nb$, acting on the particles' centres of mass.
Our analysis concerns linear responses to $\Ac_T$ and $F^\text{ext}$.
We assume that there are no particle currents in the steady state of the system, as in Ref.~\onlinecite{Wirnsberger2018}.
In response to the temperature gradient and the external force, the system will develop a density gradient parallel to $\nb$.
For convenience, we define the normalised density gradient in the same manner as the temperature gradient above, namely $\Ac_\rho=\rho^{-1}(\nb\cdot\boldnabla\rho)$.

Let $\fb_i$ be the (total) force exerted on particle $i$ by the other particles.
The equations of motion for this particle are
\begin{equation}\begin{aligned} % changed this to get a vertically centred equation number
\dot{\bm{r}}_i & = m^{-1} \bm{p}_i , & \dot{\bm{p}}_i & = \bm{F}_i , \\
\dot{\bm{e}}_i & = I^{-1} \bm{L}_i \times \bm{e}_i, & \dot{\bm{L}}_i & = \alpha( \eb_i \times \fb_i ),
\end{aligned}\label{equ:eom}\end{equation}
where $\bm{p}_i$ is the linear momentum, $\bm{L}_i$ is the angular momentum and $\bm{F}_i = F^\text{ext} \nb + \fb_i$.
The quantity $\Lb_i\cdot\eb_i$ is a constant of motion: we set $\Lb_i\cdot\eb_i=0$ for all particles.
Let 
\beq
   \rho(\bm{r}) = N\langle \updelta(\rb-\rb_i) \rangle
   \label{equ:rho}
\eeq
be the particle density at position $\rb$ in the steady state of the system.  
In this formula, the angle brackets (without any subscript) indicate an average in the steady state of the system; the (Dirac) delta  means that the mean number of particles in any spatial domain can be obtained by integrating $\rho(\rb)$ over that region.
For any single-particle observable $A$, we also define the conditional average of $A$ for particles at $\rb$.  We denote this average by angle brackets with subscript $\rb$, that is
\begin{align}
   \langle A\rangle_\rb & = \frac{ \langle A_i \updelta(\rb-\rb_i) \rangle }{ \langle \updelta(\rb-\rb_i) \rangle } 
    \nonumber \\ 
    & = \frac{N}{\rho(\rb)} \langle A_i \updelta(\rb-\rb_i) \rangle  \; ,
\end{align}
where we have used \eqnRef{equ:rho}.
A central object of interest in this work is the molecular alignment
\beq
   \langle \cos\theta\rangle_\rb =\langle \eb\cdot\nb\rangle_\rb, 
\label{equ:ave-ez}
\eeq
which measures the extent to which particle orientations are aligned with the temperature gradient or the applied body force.
% For simplicity, we sometimes refer to this quantity as the `polarisation', although there are no charges or dipoles in our molecules.
% because 'orientation' refers to the orientation vector - but it is abusing the word polarisation a little bit I think ...?

\subsection{Equations of motion for correlation functions}
 
We consider steady-state correlation functions such as $\langle A_i \updelta(\rb-\rb_i) \rangle$ and $\langle A\rangle_\rb$.
As an example, take $A_i=p_i^\mu$, which denotes a Cartesian component of the vector $\bm{p}$ (so $\mu=x,\,y,\,z$).
We compute the time derivative of the relevant expectation value and simplify it by using the equations of motion (as in Ref.~\onlinecite{Irving1950}), giving
\begin{equation}\begin{split}
   \frac{\mathrm{d}}{\mathrm{d}t} \langle p_i^\mu \updelta(\rb-\rb_i) \rangle
  & =  \langle F^\mu_i \updelta(\rb-\rb_i) \rangle
   \\ & \qquad {} + \left\langle \frac{p^\mu_i p^{\nu\vphantom{\mu}}_i}{m} \frac{\partial}{\partial r_i^\nu} \updelta(\rb-\rb_i) \right\rangle.
   \label{equ:dAdt}  % I've added a 'vphantom' to align the superscripts; otherwise they look a bit funny
\end{split}\end{equation}
Throughout this article, we use implicit summation over repeated Cartesian indices (in this case $\nu$).
The derivative of the delta function may appear problematic, but the expectation value involves an integral over $\rb_i$, so that all ambiguities can be avoided by integrating by parts.
In addition, we have
   $(\partial/\partial r_i^\nu) \updelta(\rb-\rb_i) = - (\partial/\partial r^\nu) \updelta(\rb-\rb_i)$,
and we note that the expectation value is taken at steady state, so the time derivative of the expectation value must vanish.
Hence
\beq
   0 = \langle F^\mu_i \updelta(\rb-\rb_i) \rangle - \nabla^\nu \left\langle \frac{p^\mu_i p^{\nu\vphantom{\mu}}_i}{m} \updelta(\rb-\rb_i) \right\rangle,
\label{equ:force-zero}
\eeq
where $\nabla^\nu$ indicates a derivative with respect to $r^\nu$.
Since the expectation value is a continuous function of $\rb$, this derivative exists.
This equation is a force-balance condition for particles at $\rb$.
It may be alternatively be written as
\beq
   \rho(\rb) \langle F^\mu \rangle_\rb  = \nabla^\nu \left[ \rho(\rb) \left\langle \frac{p^\mu p^\nu }{ m } \right\rangle_\rb \right],
\label{equ:force}
\eeq
which balances a body force per unit volume (left-hand side) with the divergence of the momentum flux (right-hand side).
 
\subsection{One-particle Liouville equation}

Average forces and fluxes are useful, but to obtain a more detailed analysis of the non-equilibrium steady state, we consider the probability distribution for the full state (orientation, linear momentum and angular momentum) of a particle at $\rb$,
\beq
\PP(\eb,\,\pb,\,\Lb|\rb) = \frac{N}{\rho(\rb)} \langle \updelta(\eb-\eb_i) \updelta(\pb-\pb_i) \updelta(\Lb-\Lb_i) \updelta(\rb-\rb_i) \rangle .
\eeq
The mean force acting on a particle with this state is
\beq
\bfb = \frac{ \langle \fb_i \updelta(\eb-\eb_i) \updelta(\pb-\pb_i) \updelta(\Lb-\Lb_i) \updelta(\rb-\rb_i) \rangle }
%{ \PP(\eb,\,\pb,\,\Lb|\rb)  \langle\updelta(\rb-\rb_i)\rangle}
{ \langle \updelta(\eb-\eb_i) \updelta(\pb-\pb_i) \updelta(\Lb-\Lb_i) \updelta(\rb-\rb_i) \rangle }
\eeq
which depends in general on $(\rb,\,\eb,\,\pb,\,\Lb)$.
Applying the same methodology as we did when simplifying \eqnRef{equ:dAdt}, one may enforce that the time-derivative of the steady-state distribution $\PP$ must vanish.
The result is
\begin{equation}\begin{split}
0 & =  \frac{1}{m\rho} \pb\cdot \boldnabla_\rb (\rho\PP)  + I^{-1} \boldnabla_\eb \cdot((\Lb\times \eb) \PP)
 \\ 
& \quad {} +  \boldnabla_\pb \cdot \left((\nb F^\text{ext} + \bfb)\PP\right) + \alpha\boldnabla_\Lb \cdot \left( (\eb\times\bfb) \PP \right),
\label{equ:liouville}
\end{split}\end{equation}
where we emphasise that $\rho$ depends on $\rb$, while $\PP$ and $\bfb$ both depend on $(\rb,\,\eb,\,\pb,\,\Lb)$.
This result may also be identified as the first equation of the BBGKY hierarchy.
Multiplying \eqnRef{equ:liouville} by $p^\mu$ and integrating over $\pb$, $\eb$ and $\Lb$ recovers \eqnRef{equ:force}.
In fact, \eqnRef{equ:liouville} describes the balance of all possible one-body forces and fluxes, and individual balance conditions can be obtained from it by suitable integrals.
We also emphasise that this formula is valid for positions $\rb$ in the bulk of the system, far from any reservoirs.

\section{Theory of thermally induced alignment}
\label{sec:theory}

\subsection{Motivation}

The central object of interest in this study is $\langle\cos\theta\rangle_\rb$ as defined in \eqnRef{equ:ave-ez}.
Using the notation we introduced above, the key prediction of Ref.~\onlinecite{Wirnsberger2018} for this system can then be summarised as
\begin{align}
 \langle \cos\theta\rangle_\rb & =  \frac{\alpha}{d} \left( \frac{1}{T} \left\langle \fb \cdot \nb \right\rangle_\rb - \Ac_T \right) .
\label{equ:wdfr}
\end{align}
This formula is predictive because $\Ac_T$ is the (imposed) temperature gradient, and $\langle \fb \cdot \nb \rangle_\rb$ can be computed from the equation of state (see below).
In this equation, we wrote $T$ as shorthand notation for $T(\rb)$.
We will continue to use this notation in cases where there is no ambiguity; similarly, $\rho$ will indicate $\rho(\rb)$.
\eqnRef{equ:wdfr} is given here using slightly different notation from that used in Ref.~\onlinecite{Wirnsberger2018}; we explain these differences in Appendix~\ref{app-notationComp}.

We will show how \eqnRef{equ:wdfr} can be related to the equations of motion [\eqnRef{equ:eom}] and the Liouville equation [\eqnRef{equ:liouville}].
The analysis is based on formulae that describe the extent to which equipartition of energy operates in this non-equilibrium steady state.
Note that our analysis includes the possibility of external forces that act on the particle centres of mass, so the molecular alignment can already be finite in \emph{equilibrium states}, which correspond to $\Ac_T=0$.
We return to this point below.

\subsection{Torque balance in linear response}

To obtain information about the molecular alignment, it is useful to derive a torque-balance condition that is analogous to the force-balance equation [\eqnRef{equ:force}].
One starts from an equation similar to \eqnRef{equ:dAdt}, but replacing $p_i^\mu$ by $(\Lb_i\times\eb_i)^\mu$.
The analogue of \eqnRef{equ:force} is
\beq
 \left\langle \alpha [ f^\mu - e^\mu (\eb\cdot\fb) ] - e^\mu \frac{  |\Lb|^2 }{ I }  \right\rangle_\rb
= \nabla^\nu \left\langle \frac{p^\nu (\Lb\times\eb)^\mu}{m} \right\rangle_\rb,
 \label{equ:torque-with-extra-term}
\eeq
where we used $\Lb_i\cdot\eb_i=0$ and the vector product identity $\bm{a}\times (\bm{b}\times \bm{c}) =  \bm{b}(\bm{a}\cdot\bm{c}) - \bm{c}(\bm{a}\cdot\bm{b})$. 
The expectation value on the right-hand side of \eqnRef{equ:torque-with-extra-term} vanishes in linear response, and the additional spatial gradient means that this term contributes at third order in the temperature gradient.
We therefore neglect it in the following.
This leads to
\beq
 \alpha \langle f^\mu - e^\mu f^\nu e^\nu \rangle_\rb =  \left\langle e^\mu \frac{|\Lb|^2}{I} \right\rangle_\rb  .
 \label{equ:torque}
\eeq
This equation, together with the force-balance equation [\eqnRef{equ:force}], plays a central role in what follows.

\subsection{Local equipartition (LEP) approach}
\label{sec:lep}

To make progress, we must evaluate the averages involving momentum co-ordinates in Eqs~\eqref{equ:force} and \eqref{equ:torque}.
At equilibrium ($\Ac_T=0$), this can be done by equipartition of energy.
In the linear response regime (small $\Ac_T$),
it is still consistent to use the equilibrium equipartition formula $\langle p^\nu p^\mu \rangle_\rb = m T \updelta^{\mu\nu}$ in \eqnRef{equ:force}.
(Corrections to equilibrium equipartition enter \eqnRef{equ:force} only at third order.)
Taking the scalar product with the unit vector $\nb$ yields
\beq
 \rho \Fb^\text{ext}   = \rho T (\Ac_T + \Ac_\rho ) - \rho \nb \cdot \langle \fb \rangle_\rb.
 \label{equ:force-ideal}
\eeq
We identify the ideal pressure as $P^{\rm id}=\rho T$ and the gradient of the excess pressure as $\boldnabla P^{\rm ex} = -\rho\langle \fb \rangle_\rb$.
The left-hand side of \eqnRef{equ:force-ideal} therefore corresponds to the external force per unit volume and the right-hand side corresponds to the gradient of the (total) pressure: this is mechanical equilibrium, $\rho \Fb^\text{ext} = \nabla(P^{\rm id} + P^{\rm ex})$.
If the equation of state of the fluid is known, this means that the density gradient can be computed, which enables determination of $\langle \fb\rangle_\rb$.

For systems out of equilibrium, we will show that it is not consistent to use equilibrium equipartition formulae to average the angular momenta in \eqnRef{equ:torque}.
In fact, consistency with Ref.~\onlinecite{Wirnsberger2018} requires that we take
\beq
\langle e^\mu |\Lb|^2/I\rangle_\rb = (d-1)\langle e^\mu T(\Rb_i) \rangle_\rb ,
\label{equ:mod-ep}
\eeq
where we recall that $\Rb_i$ is the position of the LJ centre of the molecule.
That is, a form of equipartition still holds, but the temperature is to be evaluated \emph{at the position of the LJ centre}, which we also identify as the `thermal centre' of the molecule.
At equilibrium, $T(\Rb_i)=T$ and one recovers the equilibrium result: the factor of $d-1$ enters because $\Lb_i\cdot\eb_i=0$, so the angular momentum has $d-1$ independent components.
A first-order Taylor expansion of the temperature in \eqnRef{equ:mod-ep} about the centre of mass gives
\beq 
\left\langle \frac{e^\mu |\Lb|^2}{ I } \right\rangle_\rb = (d-1) T \left( \left\langle e^\mu \right\rangle_\rb + \frac{\alpha \Ac_T n^\mu }{ d }  \right) .
\label{equ:emu-L2-eq}
\eeq  
The second term on the right-hand side is proportional to the distance between the molecular centre of mass and its thermal centre.
The pre-factor of this term could be altered by replacing $\Rb_i$ in \eqnRef{equ:mod-ep} by some other position within the molecule.
However, we will see later that \eqnRef{equ:mod-ep} is not arbitrary: it is constrained by the form of the Liouville equation [\eqnRef{equ:liouville}].

We assume additionally that the molecular force is uncorrelated with its orientation,
\beq
\left\langle e^\mu f^\nu e^\nu \right\rangle_\rb =  \frac{1}{ d } \left\langle f^\mu \right\rangle_\rb   ,
 \label{equ:f-indep}
\eeq
where we used $\langle e^\mu f^\nu e^\nu \rangle_\rb=\langle f^\nu \rangle_\rb \langle e^\mu e^\nu \rangle_\rb$ by independence and $\langle e^\mu e^\nu \rangle_\rb=\updelta^{\mu\nu}/d$, which is valid in linear response because the orientation is a unit vector that is  distributed isotropically at zeroth order.
At equilibrium ($\Ac_T=0$), \eqnRef{equ:f-indep} does hold; the LEP assumption is that it remains true away from equilibrium.
Combining Eqs~\eqref{equ:torque}, \eqref{equ:emu-L2-eq} and \eqref{equ:f-indep} yields \eqnRef{equ:wdfr}.

Note that if the temperature is constant throughout the system, then the assumptions of Eqs~\eqref{equ:mod-ep} and \eqref{equ:f-indep} are exact: they can be derived from the Boltzmann distribution.
In this case, \eqnRef{equ:wdfr} is exact and together with \eqnRef{equ:force-ideal} yields
$\langle \cos\theta\rangle_\rb =  \frac{\alpha}{d} \left( \Ac_\rho - F^\text{ext}/T  \right)$.
This result, which was verified numerically in the `$\nabla\rho$ runs' of Ref.~\onlinecite{Wirnsberger2018}, gives an exact prediction of particle orientation in equilibrium states with applied external forces, at the level of linear response.

\subsection{Physical interpretation of LEP}

We have derived \eqnRef{equ:wdfr} from the molecular equations of motion [\eqnRef{equ:eom}] using the assumptions of Eqs~\eqref{equ:mod-ep} and \eqref{equ:f-indep}.
This approach gives a microscopic theoretical foundation for the arguments of Ref.~\onlinecite{Wirnsberger2018}.
We will discuss these assumptions in more detail below and compare them to numerical simulations.
However, before doing so, we summarise the overall physical picture illustrated in Fig.~\ref{fig:pic}.

For thermo-orientation in the absence of an external body force, the ideal pressure gradient in \eqnRef{equ:force-ideal} is non-zero, and it must be balanced by the interparticle force $\langle\fb\rangle_\rb$.
The ideal pressure gradient is the divergence of the momentum flux: particles move more slowly in cold regions, so the time spent there tends to be longer.
The interparticle force balances out this tendency.
This force acts on the LJ centre and thus exerts a torque on the molecule (recall Fig.~\ref{fig:pic}).
Hence the force appears in the torque-balance equation [\eqnRef{equ:wdfr}].

However, molecular alignment also displaces particles' thermal centres and changes their kinetic energies.
This favours configurations where the thermal centres are closer to the cold bath, since the reduced angular velocity means that particles stay longer in these configurations.
This effect generates the temperature-gradient term in \eqnRef{equ:wdfr}.
In the example considered here, the two contributions to the right-hand side of \eqnRef{equ:wdfr} have opposite signs, with the temperature-gradient term being larger in magnitude.
This mechanical derivation and interpretation of \eqnRef{equ:wdfr} are the first key insight of this paper.

It is useful to note at this point that this LEP construction is straightforwardly extended to arbitrary shaped molecules, along the lines discussed in Ref.~\onlinecite{Gardin2019}.
The arguments of the following Section~\ref{sec:lep-dist} can also be extended, but in that case the position of the thermal centre cannot be deduced from the equations of motion, in contrast to the off-centre LJ particles considered here.

\subsection{LEP distribution}
\label{sec:lep-dist}

The assumptions of Eqs~\eqref{equ:mod-ep} and \eqref{equ:f-indep} may seem arbitrary at this stage.
We now show that these results can be justified in terms of a one-body probability distribution $\PP$ that solves the Liouville equation [\eqnRef{equ:liouville}].
That is, while alternative assumptions might appear plausible, they would not typically be consistent with \eqnRef{equ:liouville}.

We will introduce several ans\"atze for $\PP$, of the form
$$
\PP(\eb,\,\pb,\,\Lb|\rb) = \phi(\eb,\,\pb,\,\Lb|\rb) \frac{\PP_0(\eb,\,\pb,\,\Lb|\rb)}{Z},
$$
where $Z$ is a normalisation constant, $\phi$ is a smooth function and $\PP_0=\updelta(|\eb|^2-1) \updelta(\eb\cdot\Lb)$ enforces that $\eb$ is a unit vector and $\Lb$ is perpendicular to $\eb$.
%\margincomment{AR: I don't think this is particularly prone to misinterpretation, but it might perhaps be better to call generic functions something other than $f$, which we're using for the force?} yes good point.
The simplest ansatz for $\PP$ is that only the kinetic energy is correlated with the position: this motivates the name `local equipartition'.
It means that
\beq
\PP_{\rm LEP} = \frac{1}{T(\bm{r}+\alphat \eb)^{d-1/2}} \exp\mleft( \frac{-\KK}{T(\bm{r}+\alphat \eb)} + \gamma \eb\cdot\nb \mright) \frac{\PP_0}{Z_{\rm LEP}},
\label{equ:PP-LEP}
\eeq
where $\gamma$ is a parameter related to the molecular alignment, $\alphat \eb$ is the vector from the centre of mass to the thermal centre (we are treating $\alphat$ as an adjustable parameter), and
\beq
\KK = \left( \frac{|\pb|^2}{2m} + \frac{|\Lb\times\eb|^2}{2I} \right)
\label{equ:ke}
\eeq 
is the kinetic energy.  (Recall $\Lb\cdot\eb=0$, so $|\Lb\times\eb|^2=|\Lb|^2$.)  In \eqnRef{equ:PP-LEP}, $  T(\bm{r}+\alphat \eb)$ is the value of the temperature at the thermal centre.
It is shown in Appendix~\ref{app:lep-liouv} that the ansatz~\eqref{equ:PP-LEP} is consistent with \eqnRef{equ:liouville} only if the mean force on the particle is
\beq
 \bfb_{\rm LEP} = \nb \left[ u_0 T + {\Ac_T} \left( \KK - \KKb \right)  \right],
 \label{equ:bfb-LEP}
\eeq
where $\KKb=T(d-1/2)$ is the mean kinetic energy and $u_0$ is another parameter.  In addition, \eqnRef{equ:liouville} imposes
\beq
(F^\text{ext}/T) + u_0 = \Ac_T + \Ac_\rho,
\label{equ:u-f}
\eeq
which is equivalent to \eqnRef{equ:force-ideal} since $\langle\bfb\rangle=u_0\nb$, and
\beq
\gamma=\alpha(u_0-\Ac_T), \qquad \alphat=\alpha.
\label{equ:gamma-u-alpha}
\eeq
The last result for $\alphat$ means that the thermal centre in \eqnRef{equ:PP-LEP} must indeed coincide with the LJ centre, as we claimed above.

Evaluating averages with respect to $\PP_{\rm LEP}$, one finds in linear response that
$$
\langle \cos\theta\rangle_\rb = \gamma/d \qquad\text{and} \qquad \langle \fb \rangle_\rb = \nb u_0 ,
$$
which together with \eqnRef{equ:gamma-u-alpha} imply \eqnRef{equ:wdfr}.
The other LEP assumptions [Eqs~\eqref{equ:mod-ep} and \eqref{equ:f-indep}] can also be derived as averages with respect to $\PP_{\rm LEP}$.  Also, similarly to \eqnRef{equ:emu-L2-eq}, we have
\beq 
\left\langle \frac{e^\mu |\pb|^2}{ m } \right\rangle_\rb = Td \left( \left\langle e^\mu \right\rangle_\rb + \frac{\alpha \Ac_T n^\mu }{ d }  \right) .
\label{equ:emu-p2-eq}
\eeq  

The conclusion of this analysis is that the LEP distribution $\PP_{\rm LEP}$ and its associated force $\bfb_{\rm LEP}$ describe a consistent physical picture, at the one-body level, of the behaviour of molecules in this system.
If the equation of state of the fluid is given, then the distribution has no adjustable parameters.
This picture is fully consistent with Ref.~\onlinecite{Wirnsberger2018} and has the same implications.
However, the LEP distribution also makes predictions for other correlation functions, beyond those considered so far.

In fact, our numerical results (Sec.~\ref{sec:numerics})  show that the LEP distribution does not provide a full description of the linear response of this fluid.
To this end, we discuss some corrections to this distribution.

An important feature is that the LEP distribution is time-reversal symmetric (i.e.~it is invariant under reversal of all momenta), so it cannot describe dissipative effects such as heat currents, which will be present if $\Ac_T\neq0$.
Dissipative effects lead to non-zero values of correlation functions that are odd under time reversal, such as $\langle p^\mu |\pb|^2/m\rangle$.
\eqnRef{equ:PP-LEP} can be modified in order to incorporate such correlations, but within linear response, these modifications have no effect on non-dissipative (time-reversal-symmetric) quantities such as the molecular alignment.
We therefore neglect dissipative terms in the following.
However, there are also {corrections} to LEP that do affect the molecular alignment, which we discuss in Sec.~\ref{sec:epb}.

\subsection{Equipartition breaking (EPB)}
\label{sec:epb}

We consider alternative solutions to \eqnRef{equ:liouville}, which we write in terms of a function $\updelta G$ as
\beq
\PP_{\rm EPB} = \PP_{\rm LEP} ( 1 + \alpha\Ac_T \updelta G ).
\label{equ:pp-delta}
\eeq
Here EPB indicates an `equipartition-breaking' solution (and note that $\delta G$ indicates a change in $G$, the $\delta$ does not indicate any kind of delta function).  
Similarly
\beq
 \bfb_{\rm EPB} = \bfb_{\rm LEP} + \alpha\Ac_T (\updelta\bfb).
 \label{equ:bfb-delta}
\eeq
The factors of $\alpha\Ac_T$ in these equations highlight that corrections to LEP are assumed to be linear in the deviation from equilibrium, and are odd in $\alpha$.
{Expansions similar to Eq.~(\ref{equ:pp-delta}) have been considered before in kinetic theory~\cite{chapman-book} and also (for example) when analysing systems where energy is slowly injected into a homogeneous system, during a chemical reaction}\cite{Groot1984,Ross1961}.  {In general, one expects that the Maxwell-Boltzmann distribution for $\bm{p}$ can be perturbed by the thermal gradient, and also that the internal degrees of freedom of the molecule (orientation $\eb$ and angular momentum $\Lb$) can become correlated with $\pb$, and with each other.
In the situation considered here, we emphasize that $\delta G$ must be odd in $\bm{n}$: the terms that appear at first order in our expansion change sign if the temperature gradient is reversed.  This symmetry greatly simplifies the analysis by restricting the terms that can appear in our ansatze for $\updelta G$ and $\updelta\bfb$.  One possible correlation, consistent with this symmetry, is that particles whose orientations are aligned with the temperature gradient ($\eb\cdot\nb>0$) have slightly higher kinetic energies, compared with particles that are anti-aligned ($\eb\cdot\nb<0$). See Eqs.~(\ref{equ:EEL},\ref{equ:EEP}) below.}

We have found a three-parameter family of EPB solutions, which depend on (dimensionless) free parameters $\kappa$, $\lambda$ and $\xi$.
We note that this solution is specific to this system, in which all intermolecular forces act via a single `force centre'.
This means that the only difference between the forces $\Fb$ and $\fb$ in \eqnRef{equ:eom} is the external force.
This places strong constraints on the possible solutions to \eqnRef{equ:liouville}, and enables this analysis.
The formulae that describe the EPB solutions are somewhat unwieldy: we first state them and then discuss their physical consequences.

The EPB distribution has
\begin{multline}
\updelta G = - \frac{\nb\cdot\eb}{\chi T} \left(  \frac{\pb}{m} +\alpha\dot\eb \right) \cdot \left( \kappat \pb + \lambdat m \alpha \dot\eb  \right)
\\ + \xi \frac{\pb\times\Lb\cdot\nb}{m\alpha T} 
+ (\nb\cdot\eb) \left[ \kappat - (\kappat d/\chi) - \lambdat d \right]
\label{equ:delta-P}
\end{multline}
and the associated mean force is
\begin{multline}
\updelta\bfb= 
\eb(\nb\cdot\eb) (\lambdat + \kappat) \frac{|\Lb|^2}{I\alpha} 
-
\frac{ \nb\cdot \dot\eb  }{\chi }
\left( \kappat\pb + \lambdat m \alpha \dot\eb \right) 
\\ 
- \frac{\kappat T}{\alpha} \left[ \nb - \eb(\nb\cdot\eb)  \right] ,
\label{equ:bfb-epb}
\end{multline}
where  
%(\emph{notation\dots  maybe put a factor of $\alpha\Ac_T$ in front of everything to emphasise that we are in linear response?}), 
we recall $\chi=m\alpha^2/I$.
%(\emph{should flip signs so that $\kappa,\lambda$ end up positive.  At some places I have used notation $\kappat=-\kappa$ and $\lambdat=-\lambda$.})
The consistency of these formulae with \eqnRef{equ:liouville} is demonstrated in Appendix~\ref{app:epb-liouv}.

The parameters $\kappa$, $\lambda$ and $\xi$ are not predicted within this theory, so this general form for $\PP$ does not allow predictions from first principles, contrary to LEP.
However, the EPB theory does enforce relations between correlation functions that come from the equations of motion, such as \eqnRef{equ:torque}.
Hence, if $\kappat$, $\lambdat$ and $\xi$ are obtained by measuring certain correlation functions, then the values of other correlation functions can be predicted.
If all these parameters are zero, then we recover LEP.

For the cases considered below, $\kappat$, $\lambdat$ and $\xi$ are all positive and of similar magnitudes.
The physical implications of these parameters become apparent when we compute correlation functions (see also Appendix~\ref{app:correl}). 
For example, instead of \eqnRef{equ:emu-L2-eq}, we find
\beq 
\frac{1}{T(d-1)} \left\langle e^\mu \frac{|\Lb|^2}{I} \right\rangle_\rb -  \langle e^\mu \rangle_\rb = \frac{\alpha \Ac_T n^\mu}{d} ( 1 - 2\lambdat ) .
\label{equ:EEL}
\eeq 
The LEP case corresponds to $\lambdat=0$.
We see that {positive $\lambdat$} corresponds to a weaker correlation between a particle's angular momentum and its orientation compared with LEP.
Similarly
\beq
\frac{1}{Td} \left\langle e^\mu \frac{|\pb|^2}{m} \right\rangle_\rb -  \langle e^\mu \rangle_\rb = \frac{\alpha \Ac_T n^\mu}{d} ( 1 - 2\kappat/\chi ) .
\label{equ:EEP}
\eeq
Based on these equations, one possible physical interpretation is that the thermal centre of the molecule is displaced away from the LJ centre towards the centre of mass.
However, we argue that this is not appropriate: a key feature of EPB is that one cannot use a single thermal centre to account for the statistics of both linear and angular momentum.

%(\ref{equ:f-indep}),
As well as modified equipartition formulae, EPB also predicts correlations of molecular properties with the force.  For example, \eqnRef{equ:f-indep} becomes
\beq
T^{-1} \left\langle \frac{f^\mu}{d}  - e^\mu f^\nu e^\nu \right\rangle_\rb  = -\Ac_T n^\mu (\lambdat+\kappat)\frac{(d-1)}{d}.
\label{equ:Qpar}
\eeq
The LEP case is $\lambdat=\kappat=0$, in which case the left-hand side is zero, consistent with \eqnRef{equ:f-indep}.
The EPB distribution predicts that the force is correlated with the molecular orientation.
In particular, for $\lambdat+\kappat>0$, it predicts that the intermolecular force tends to be larger when the molecules are either parallel or antiparallel to the temperature gradient.
This effect tends to reduce the torque on the particles.

Combining the torque-balance condition [\eqnRef{equ:torque}] with Eqs~\eqref{equ:EEL} and \eqref{equ:Qpar} leads to a modified prediction for the molecular alignment,
\beq
 \langle \cos\theta\rangle_\rb  =  \frac{\alpha}{d} \left( \frac{1}{T} \left\langle \fb \cdot \nb \right\rangle_\rb - \Ac_T(1-\lambdat+\kappat) \right)  ,
 \label{equ:pol-epb}
\eeq
which generalises \eqnRef{equ:wdfr}.
% We test this equation below in the light of numerical data.
Note that if for some reason we have $\lambda \approx \kappa$, then \eqnRef{equ:wdfr} may still hold to high accuracy, even if the LEP assumptions [Eqs~\eqref{equ:mod-ep} and \eqref{equ:f-indep}] have significant violations.

We have seen that $\kappa$ and $\lambda$ are related to deviations from equipartition, and also have implications for correlations between force and orientation.
The third parameter $\xi$ is related to coupling between linear and angular momentum.
For example,  one has
\beq
 \frac{1}{IT} \langle \pb\times\Lb\cdot\nb \rangle_\rb = \Ac_T(2\xi-\lambdat-\kappat)\frac{d-1}{d}  .
\label{equ:D-cor}
\eeq
Another quantity related to equipartition is
\beq
\frac{1}{T} \langle e^\mu (\pb\cdot\dot\eb) \rangle_\rb =\Ac_T n^\mu (\xi-\lambdat-\kappat)\frac{d-1}{d}  .
\label{equ:Etimes}
\eeq
One notes that among the five correlation functions [Eqs~\eqref{equ:EEL}--\eqref{equ:Qpar}, \eqref{equ:D-cor} and \eqref{equ:Etimes}],  %(Eqs~\eqref{equ:EEL}, \eqref{equ:EEP}, \eqref{equ:Qpar}, \eqref{equ:D-cor} and \eqref{equ:Etimes}],
there are only three independent parameters, $\kappa$, $\lambda$ and $\xi$.
These constraints on correlation functions arise from the form of the Liouville equation [\eqnRef{equ:liouville}].   For example, in Appendix~\ref{app:correl}, \eqnRef{equ:epb-triv2} is derived directly from the equations of motion (in linear response).  It can be rearranged to give
\beq
\frac{1}{T} \langle e^\mu (\pb\cdot\dot\eb) \rangle_\rb =  \frac{1}{2IT} \langle (\pb\times\Lb)^\mu \rangle_\rb - \frac{1}{2T} \langle (f^\mu/d) - e^\mu (\fb\cdot\eb) \rangle_\rb
\eeq
This result must hold for any solution of the Liouville equation, which of course includes the EPB solution.
Note however that \eqnRef{equ:Qpar} cannot to our knowledge be derived directly from the equations of motion: it seems to be a specific property of the EPB solution.

\subsection{Physical interpretation of EPB}

Physically, we interpret $\lambdat$ and $\kappat$ in terms of the correlations between orientation and the kinetic energy of a particle, from Eqs~\eqref{equ:EEL} and \eqref{equ:EEP}.
If $\lambdat>0$ and $\kappat>0$, the kinetic energy of a particle is less strongly correlated with its orientation than is predicted by LEP.
Similar effects could also be achieved by assuming that the thermal centre of the molecule is somewhere between the LJ centre and the molecular centre of mass, instead of at the LJ centre, as in LEP.
However, the form of \eqnRef{equ:delta-P} and the fact that the correlation function in \eqnRef{equ:Etimes} is non-zero within EPB both show that the state described by \eqnRef{equ:delta-P} cannot be accounted for by assuming the existence of a single thermal centre.  

In numerical simulation (see Sec.~\ref{sec:numerics}), the parameters $\kappat$ and $\lambdat$ can be estimated by considering correlations of orientation and velocity.
This leads to non-trivial predictions for the correlations between the interparticle force and the velocity, via \eqnRef{equ:bfb-epb}.
In particular, one sees from \eqnRef{equ:Qpar} that the correlations between velocities and orientations can only be sustained if there are also correlations of the interparticle force with the orientation.

The parameter $\xi$ is different in that it does not appear in the expression for the mean force [\eqnRef{equ:bfb-epb}].
In this sense, it encodes correlations among particle velocities that do not require any (mean) force to sustain them.
Our numerical results (see below) are consistent with $\lambdat$, $\kappat$ and $\xi$ all being of comparable magnitudes, with typical values in the range $0.1$--$0.2$.

\section{Numerical results}\label{sec:numerics}

\subsection{Calculation method}

We obtained numerical results using  a modified version of the LAMMPS simulation package~\cite{Plimpton1995} (v.~11Aug17), with the same methods as~Ref.~\onlinecite{Wirnsberger2018}.
The system is periodic, with box dimensions $L\times L\times 2L$, and comprises 5832 off-centre LJ particles.
The box length is chosen to produce the desired overall number density, and in our simulations is of the order of $L\approx 15\sigma$.
The LJ interaction is truncated at $7\sigma$ and the time step was $0.005 \sigma \sqrt{m/\varepsilon}$.
There is a temperature gradient along the $z$ direction, which is achieved by imposing different temperatures in two equally spaced thermal baths, each of width $2\sigma$.
The temperature in these regions is controlled using 
 local Gaussian thermostats that act on the non-translational kinetic energy of the reservoir and leave the centre of mass motion unaffected,
%Gaussian thermostats, 
as in Ref.~\onlinecite{Wirnsberger2018}.  After equilibration at an appropriate state point~\cite{Wirnsberger2018}, the Gaussian thermostats were activated and the system was simulated for a time $2\times 10^3 \sqrt{m\sigma^2/\varepsilon}$, in order to establish the non-equilibrium steady state.  This was followed by production runs of $3.45\times 10^5\sqrt{m\sigma^2/\varepsilon}$.  (These long simulations are necessary because the magnitude of the thermo-orientation effect is small, so significant averaging is required in order to obtain small statistical uncertainties.)
%$6.9\times 10^7$ time steps, 
%.

We partitioned the system into 20 segments along the $z$ direction and computed single-particle observables of interest independently within each segment.
The results shown here are averaged over the segments, but excluding regions that are overlapping with or close to the thermal baths.
The dependence of the results on $z$ is very weak in all cases, consistent with the small imposed temperature gradient.
Error bars are computed by assuming that measurements within each segment are independent and calculating the standard error.
Similar error estimates could also be obtained by analysing how the results fluctuate as a function of time.  
In addition to the results shown here, we also carried out some simulations with a smaller time step $0.002 \sigma \sqrt{m/\varepsilon}$. 
These simulations reached shorter times so there is less data to average over, hence numerical uncertainties are larger.
However, we did not find any evidence for significant dependence of our results on the timestep.

\subsection{Dependence of orientation on $(\rho,\,T)$}

\begin{figure}
\includegraphics{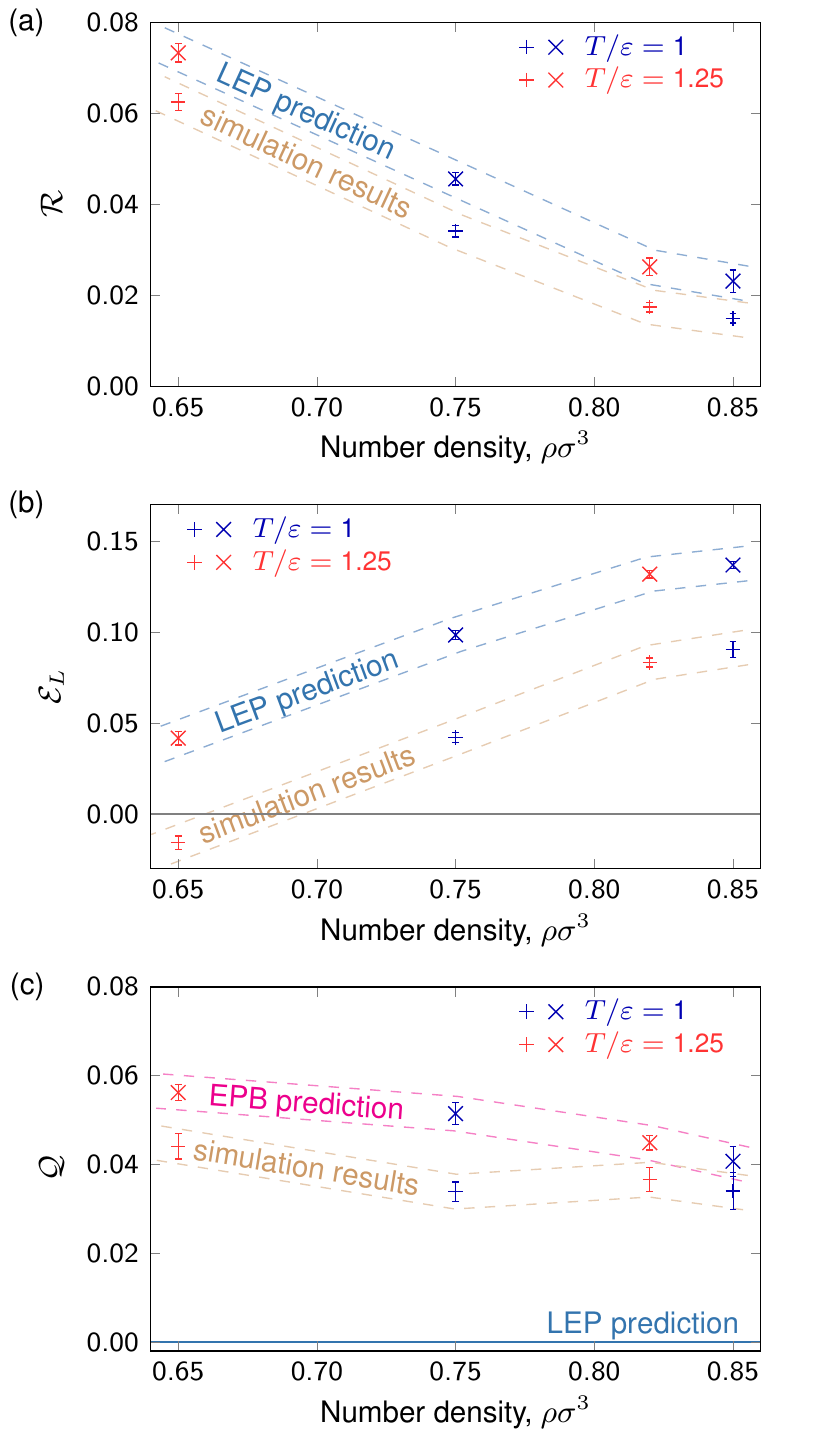}
\caption{(a)~Orientational response $\cal R$, shown as a function of density $\rho$, for two different temperatures.  The results are compared with the LEP prediction [\eqnRef{equ:R-lep}, equivalent to \eqnRef{equ:wdfr}], showing good, but not perfect, agreement.  (The data are labelled by the average temperature, dashed lines are guides to the eye.)
(b) The coupling ${\cal E}_L$ of rotational kinetic energy with orientation.  The LEP prediction [\eqnRef{equ:EL-lep}, equivalent to \eqnRef{equ:mod-ep}], is considerably worse in this case.
(c) The correlation ${\cal Q}$ of force with orientation, which is predicted to be zero within LEP, see \eqnRef{equ:Q-lep}.  The LEP prediction is poor. The EPB prediction [\eqnRef{equ:calQ-epb}, equivalent to \eqnRef{equ:Qpar}] for this quantity is significantly better, although still not perfect.}
\label{fig:state}
\end{figure}

This section shows results for several state points, which are labelled by (average) temperatures $T=1.0\eps$ and $T=1.25\eps$.  The heat baths are maintained at $0.95T$ and $1.05T$.  The local temperature in simulations is determined as $\langle |\pb|^2/(md)\rangle_\rb$: this varies linearly with $z$ in the region outside the baths.  The temperature gradient $\nabla T$ is measured using data for the local temperature, in the region outside of the baths.  This leads to values of $\Ac_T$ between $0.006/\sigma$ and $0.008/\sigma$.

Since these gradients are small, the responses are also small, and we therefore normalise all responses by the size of the gradient itself.
To this end, we define dimensionless observables, which all have the property that they are independent of $\Ac_T$ within the linear response regime.
The response of the orientation to the gradient is
\begin{equation}
{\cal R} = \frac{1}{\sigma\Ac_T} \langle \cos\theta \rangle_\rb \;.
\label{equ:RF}
\end{equation}
The normalised interparticle force is
\begin{align}
{\cal F} & = \frac{1}{T\Ac_T} \langle f^z \rangle_\rb.
\end{align}
With this choice, and since $\nb$ is a unit vector in the $z$ direction, the LEP prediction [\eqnRef{equ:wdfr}] for the alignment, which is also the prediction of Ref.~\onlinecite{Wirnsberger2018}, is
\beq
{\cal R} = \frac{-\alpha}{\sigma d} ( 1- {\cal F}  )  \;.
\label{equ:R-lep}
\eeq
Note that $\alpha<0$ throughout our numerical work.  In this section, we take $\alpha=-0.25\sigma$, as in Ref.~\onlinecite{Wirnsberger2018}.
The LEP prediction of \eqnRef{equ:R-lep} is tested in Fig.~\figrefsub{fig:state}{(a)}, for several values of $(\rho,\,T)$.
There is good, but not perfect, agreement between the prediction and the numerical results.
The deviations are comparable with those found in the NEMD (non-equilibrium molecular dynamics) simulations of  Ref.~\onlinecite{Wirnsberger2018}.

It is also notable that the response $\cal R$ decreases with density.  This is expected because Eqs~\eqref{equ:wdfr} and \eqref{equ:force-ideal} with $F^\text{ext}=0$ together imply ${\cal R} = \Ac_\rho/(\Ac_Td)$, so the response is proportional to the induced density gradient $\Ac_\rho$.  As the fluid gets denser, the compressibility is reduced, so $\Ac_\rho/\Ac_T$ decreases, and so does the thermo-orientation effect.

To test the LEP theory in more detail, we define two quantities that measure the correlation between orientation and kinetic energy,
\begin{align}
{\cal E}_L & = \frac{-1}{\sigma T\Ac_T} \left\langle e^z \frac{|\Lb|^2}{I} \right\rangle_\rb \;,
 \\
{\cal E}_p & = \frac{-1}{\sigma T\Ac_T} \left\langle e^z \frac{|\pb|^2}{m} \right\rangle_\rb \;.
\end{align}
The minus signs in these definitions are somewhat arbitrary: they are included so that ${\cal E}_L$ and ${\cal E}_p$ are positive in our numerical calculations, which means that the orientation is anti-correlated with the kinetic energy.
We also define
\beq
{\cal Q} = \frac{-\alpha}{\sigma T\Ac_T} \left\langle e^z (\fb\cdot\eb) - (f^z/d) \right\rangle_\rb,
\label{equ:def-Q}
\eeq
which measures the correlation between the interparticle force and the orientation.
The LEP predictions [Eqs~\eqref{equ:emu-L2-eq} and \eqref{equ:f-indep}] are that
\begin{align}
{\cal E}_L  & = \left(  \frac{-\alpha}{\sigma d} - {\cal R} \right) (d-1) ,
\label{equ:EL-lep}
\\
{\cal Q} & = 0 .
\label{equ:Q-lep}
\end{align}
These two predictions are tested in Fig.~\figrefsub{fig:state}{(b,\,c)}.  One sees significant violations of these predictions, which are larger than the deviations from \eqnRef{equ:R-lep} that are apparent in Fig.~\figrefsub{fig:state}{(a)}.
In Sec.~\ref{sec:lep}, assumptions equivalent to Eqs~\eqref{equ:EL-lep} and \eqref{equ:Q-lep} were used to derive the result given by \eqnRef{equ:wdfr}, which is equivalent to \eqnRef{equ:R-lep}.
The results of Fig.~\ref{fig:state} show that the final result holds more accurately than the assumptions that were used to derive it.
Clearly, some cancellation of errors is at work.

To investigate this, we consider the EPB theory, which has adjustable parameters $\kappat$ and $\lambdat$.
To estimate them, we use the EPB predictions of Eqs~\eqref{equ:EEL} and \eqref{equ:EEP} to write
\begin{equation}\begin{split}
\lambdat & = \frac12 \left[ 1 + \frac{\sigma d}{\alpha} \left( {\cal R} + \frac{{\cal E}_L}{d-1} \right) \right]
%\frac{d}{(d-1)} \left( 1 + \frac{ {\cal R} (d-1) + {\cal E}_L}{\alpha/\sigma} \right),
\\
\kappat & = \frac{\chi}{2} \left[ 1 + \frac{\sigma d}{\alpha} \left(  {\cal R} + \frac{{\cal E}_p}{d} \right) \right].
\label{equ:lambda-kappa-num}
\end{split}
\end{equation}
Hence $(\kappat,\,\lambdat)$ may be estimated from the data.  (If LEP holds, then $\kappat=0=\lambdat$.)
Once $\kappat$ and $\lambdat$ are fixed, then \eqnRef{equ:Qpar} is a non-trivial prediction of EPB theory.
It implies that
\beq
{\cal Q} = \frac{-\alpha}{\sigma} (\lambdat + \kappat )  \frac{d-1}{d}
\label{equ:calQ-epb}
\eeq
This prediction is tested in Fig.~\figrefsub{fig:state}{(c)}.
We note that in contrast to LEP, which predicts ${\cal Q}=0$, the EPB prediction gives the correct (positive) sign for $\cal Q$.
However, the numerical results still deviate from the EPB prediction.
Recalling \eqnRef{equ:def-Q}, we observe from Fig.~\figrefsub{fig:state}{(c)} that the force $\fb$ tends to be either parallel or anti-parallel to $\eb$ (recall $\alpha<0$).  Hence, the mean torque on the molecule is reduced, compared to the LEP theory.
This situation is similar for all four state points considered.
We now fix a single state point and study its behaviour in  more detail, focussing on the dependence of thermo-orientation on $\alpha$.

\begin{figure}
\includegraphics{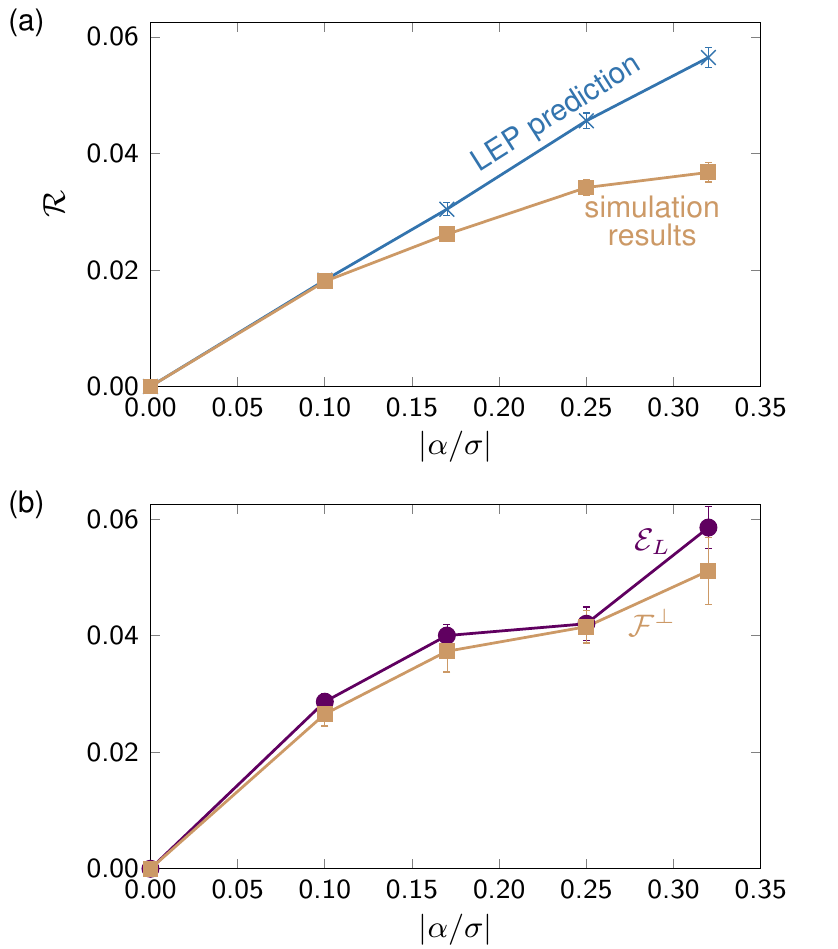}
\caption{(a)~Orientational response $\cal R$ as a function of $|\alpha/\sigma|$, and the LEP prediction [\eqnRef{equ:R-lep}].  The agreement is good for small $|\alpha|$ but deviations appear for larger $|\alpha|$. (b)  Test of the torque-balance condition [\eqnRef{equ:EL-torque}], which is derived directly from the equations of motion.  This is obeyed to within our numerical uncertainty.}
\label{fig:data-ez}
\end{figure}

\begin{figure}
\includegraphics{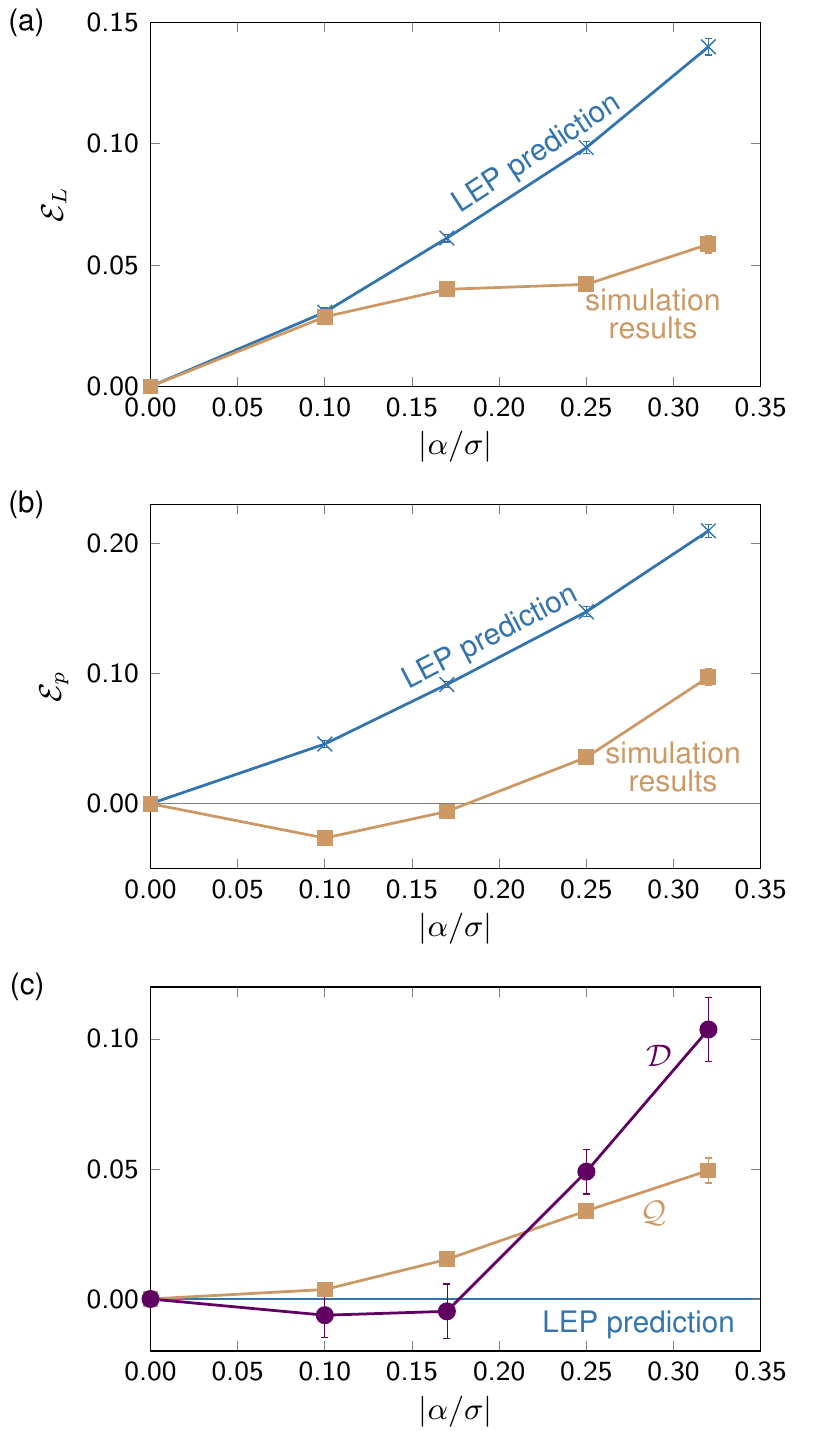}
\caption{Deviations from LEP. (a,b) Results for the coupling between orientation and kinetic energy, ${\cal E}_L$ and ${\cal E}_p$, compared with the LEP predictions [Eqs~\eqref{equ:EL-lep} and \eqref{equ:EP-lep}].
There are significant deviations from LEP.
(c) The correlation $\cal Q$ between force and orientation and the correlation $\cal D$ between translational and angular momentum, both of which are predicted to be zero in LEP.
}
\label{fig:lep-dev}
\end{figure}

\begin{figure}
\includegraphics{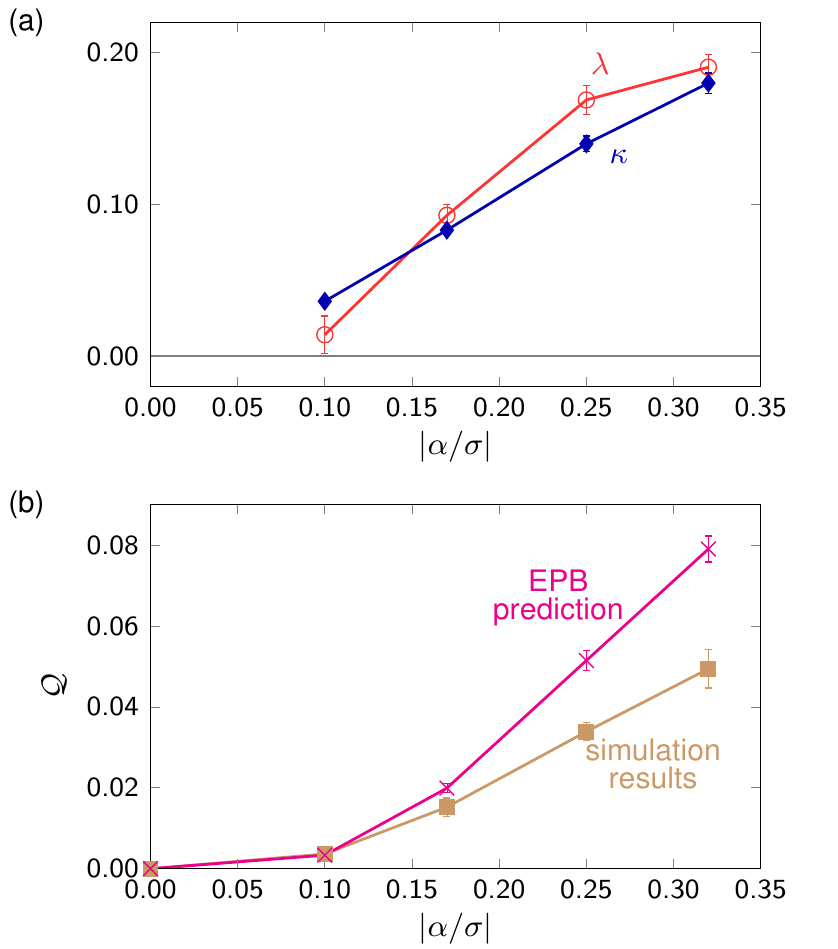}
\caption{EPB theory.  (a) Values of the EPB parameters $\kappat,\lambdat$, estimated using \eqnRef{equ:lambda-kappa-num}.
(b) Test of the EPB prediction [\eqnRef{equ:Qpar}], based on the values of $\kappat,\lambdat$ in the top panel.  The EPB theory partially captures the deviations from LEP, but it predicts a stronger correlation than is found numerically.}
\label{fig:epb}
\end{figure}

\subsection{Dependence of orientation on $\alpha$}

We focus on $T=1.0\eps$ and $\rho=0.75\sigma^{-3}$, and we consider how the results depend on $\alpha$.  For $\alpha=0$, there is no thermally induced alignment, so it is natural to expect responses proportional to $\alpha$ when $|\alpha/\sigma|\ll 1$.
However, we emphasise that there is no assumption of small $\alpha$ within our theory: we consider linear response to the thermal gradient $\Ac_T$, but our theory can in principle capture non-linear dependence on $\alpha$.

Fig.~\figrefsub{fig:data-ez}{(a)} shows the response as a function of $\alpha$, and the comparison with the LEP prediction of \eqnRef{equ:wdfr}.
[The points for $\alpha=-0.25\sigma$ were already shown in  Fig.~\figrefsub{fig:state}{(a)}.]
The LEP prediction is almost linear in $\alpha$, because the mean force $\langle f \rangle_\rb$ in \eqnRef{equ:wdfr} depends very weakly on $\alpha$.
The numerical results show a saturation in the thermo-orientation as $\alpha$ is increased, which is not captured by \eqnRef{equ:wdfr}.
We have checked that this result is \emph{not} associated with a breakdown of linear response in $\Ac_T$: for $\alpha=-0.32\sigma$, we reduced the temperature gradient by a factor of two, which led to very small change in $\cal R$ (the difference was less than $10^{-3}$ and within the range of error bars).
The conclusion is that LEP is violated at the level of linear response to the temperature gradient.
This is consistent with the EPB theory.

We now consider the torque-balance equation [\eqnRef{equ:torque}].
We define a normalised torque as the projection of the interparticle force perpendicular to the orientation,
\beq
{\cal F}^\perp = \frac{-\alpha}{\sigma T \Ac_T} \langle f^z - e^z (\fb\cdot\eb) \rangle_\rb \;,
\eeq
so that \eqnRef{equ:torque} is equivalent to
%In Fig.~\ref{fig:data-ez}(b), we show the prediction of the torque-balance equation (\ref{equ:torque}), which in the notation of this section is
\beq
{\cal E}_L =  {\cal F}^\perp.
\label{equ:EL-torque}
\eeq
From Fig.~\figrefsub{fig:data-ez}{(b)}, we see that the equation is obeyed, up to numerical and statistical uncertainties.
It was derived directly from the equations of motion, up to our neglect of the right-hand side of \eqnRef{equ:torque-with-extra-term}, which we have separately verified to be very small [at least one order of magnitude smaller than the quantities plotted in Fig.~\figrefsub{fig:data-ez}{(b)}].
Since \eqnRef{equ:torque} does not rely on LEP nor on linear response, any deviations from this prediction would have to be attributed to errors associated with our numerical integration of the equations of motion.

We now test in detail the predictions of LEP, which include Eqs~\eqref{equ:EL-lep} and \eqref{equ:Q-lep}, as well as \eqnRef{equ:emu-p2-eq}, which is equivalent to
\beq
{\cal E}_p =  \frac{-\alpha}{\sigma} -{\cal R} d.
\label{equ:EP-lep}
\eeq
We also define
\beq
{\cal D} = \frac{1}{IT\Ac_T} \langle (\pb \times \Lb)^z \rangle
\eeq
and note that LEP  predicts 
\beq
{\cal D} = 0.
\label{equ:D-lep}
\eeq
The predictions of Eqs~\eqref{equ:EL-lep}, \eqref{equ:Q-lep}, \eqref{equ:EP-lep} and \eqref{equ:D-lep} are all tested in Fig.~\ref{fig:lep-dev}. It is interesting to note that while the LEP prediction in Fig.~\figrefsub{fig:data-ez}{(a)} is accurate for $|\alpha/\sigma|=0.17$ and gives a reasonable estimate of the orientation effect at $|\alpha/\sigma|=0.25$, the other LEP predictions shown in Fig.~\ref{fig:lep-dev} are much less accurate in this regime.   This is a similar cancellation of errors to that observed in Fig.~\ref{fig:state}: that is, LEP gives reasonable predictions of $\cal R$, but less accurate predictions of other quantities.

The EPB theory provides a partial explanation of this behaviour.
The theory predicts \eqnRef{equ:pol-epb}, which is equivalent to
\beq
{\cal R} = \frac{-\alpha}{\sigma d} ( 1- {\cal F}  + \kappat - \lambdat ) .
\label{equ:R-epb}
\eeq
We can compare this expression with \eqnRef{equ:R-lep}: the difference is in the term $\kappat - \lambdat$.
Fig.~\figrefsub{fig:epb}{(a)} shows estimates of $\kappat$ and $\lambdat$ using \eqnRef{equ:lambda-kappa-num}.  Both are positive and significantly different from zero, consistent with the deviations from LEP seen in Fig.~\figrefsub{fig:lep-dev}{(a,\,b)}.   Note also that our results are consistent with $\kappat\to 0$ and $\lambdat\to0$ as $\alpha\to0$, so that LEP appears to hold for small $\alpha$.

The key point is that $\kappat$ and $\lambdat$ are numerically similar, which means that the EPB contributions to \eqnRef{equ:R-epb} are small, even though the individual values of $\kappat$ and $\lambdat$ indicate significant violations of LEP.
Results are shown in Fig.~\figrefsub{fig:epb}{(b)}, compared with the EPB prediction [\eqnRef{equ:calQ-epb}].
For $\alpha=-0.17\sigma$, our results are consistent with EPB theory: we have $\kappat\approx\lambdat$ and the LEP prediction for ${\cal R}$ is still accurate, even if other LEP predictions fail.
For $|\alpha/\sigma|\geq 0.25$, the EPB theory still gives better qualitative predictions than LEP, but we also observe significant deviations, consistent with Fig.~\figrefsub{fig:state}{(c)}.
While the LEP theory predicts ${\cal Q}=0$, the EPB theory gives the right sign for $\cal Q$ and is accurate for small $|\alpha/\sigma|$, but there are deviations for larger values of $|\alpha/\sigma|$.  

Finally, we note from \eqnRef{equ:D-cor} that
\beq
{\cal D} = (2\xi - \lambdat - \kappat) \frac{d-1}{d}.
\eeq
For $|\alpha/\sigma|<0.25$, where the EPB theory is reasonably accurate, one may then use the results of Fig.~\ref{fig:epb} to estimate $\xi$, which is positive and similar in magnitude to both $\kappat$ and $\lambdat$.  Within the framework presented here, we have no explanation of this apparent coincidence.  The smallness of $\kappat-\lambdat$ is the source of the cancellation of errors in the LEP prediction of $\cal R$, so it would be useful to have an explanation of this effect.  It is notable that for $\kappat=\lambdat$, the first term in \eqnRef{equ:delta-P} is proportional to the squared velocity of the LJ centre of the molecule, but a more detailed analysis of the structure of $\delta G$ is beyond the scope of this work.

\section{Conclusion}
\label{sec:conc}

We have shown how a theory of thermo-orientation can be derived directly from the equations of molecular dynamics.  The central idea is that the form of the one-particle Liouville equation [\eqnRef{equ:liouville}] places constraints on steady-state correlation functions of the non-equilibrium fluid.  The mean force $\bfb$ in that equation depends on the orientation and velocity of the particle, and this dependence is not known.  However, we have presented two possible types of solution for \eqnRef{equ:liouville}, which are associated with specific forms for $\bfb$.  One of these solutions corresponds to a local equipartition assumption (LEP) and the other to a non-trivial state where local equipartition is broken (EPB).   The LEP assumption recovers the prediction of Ref.~\onlinecite{Wirnsberger2018} for thermo-orientation in this system.  The EPB assumption shows how deviations from LEP can appear even at linear response in the temperature gradient.

We have also presented numerical results.  For molecules where the centre of mass is close to the interaction centre (small $|\alpha/\sigma|$), the LEP theory is very accurate.  For $|\alpha/\sigma|\gtrsim 0.25$, there are significant deviations from LEP.  The EPB theory can capture some of these deviations by appropriate choices of fitting parameters.   This theory also makes a non-trivial prediction [\eqnRef{equ:calQ-epb}, equivalent to \eqnRef{equ:Qpar}] which improves on the LEP prediction but is still not fully consistent with the numerical results.  This indicates that the EPB theory is not a full description of the non-equilibrium steady state, even in the linear-response regime.

The detailed EPB theory is restricted to the system considered here, because specifying the mean force $\bfb$ in the Liouville equation means that the mean torque on the molecule is also fully determined.  However, the structure of the LEP solution is quite general, and can be analysed in other cases too.  For future directions, one might include electrostatic polarisation in the theory~\cite{Wirnsberger2018} as well as investigating in more detail the accuracy of LEP for particles with different shapes~\cite{Gardin2019}.

{Another possible direction would be to connect with classical kinetic theory calculations that start from the Boltzmann equation}~\cite{chapman-book,Ross1961}.
{This might lead to additional constraints on the terms appearing in the expansion (\ref{equ:bfb-delta}) for $\overline{\bm{f}}$, as well as helping to relate these terms to microscopic physical processes.  The link to kinetic theory also suggests that one might investigate thermo-orientation in fluids that are more dilute, instead of dense liquids.  For example, an expansion in powers of the density might offer a systematic characterisation of the possible orientational responses to temperature gradients, and their expected magnitudes.}

\begin{acknowledgments}
We thank Daan Frenkel, Christoph Dellago, Kranthi Mandadapu, and Ronojoy Adhikari for helpful discussions.  P.W. is grateful for a Microsoft Azure Research Sponsorship.  The results presented here were achieved in part using the Vienna Scientific Cluster.
{The data generated during this study are available at {https://doi.org/10.17863/CAM.37306}.}
\end{acknowledgments}

\begin{appendix}

\section{Connection to previous studies}\label{app-notationComp}

In Eq.~(5) of Ref.~\onlinecite{Wirnsberger2018}, the alignment of non-polar molecules is computed via
\begin{equation}
\langle \cos \theta(z) \rangle = \frac{\beta(z)\alpha f_\text{LJ}(z)}{3} = \frac{\alpha f_\text{LJ}(z)}{d T(z)} ,
\label{equ:wdfr-compare}
\end{equation}
where $f_\text{LJ}$ is a force that acts at the LJ centre of the molecule, and we have used $d=3$ explicitly.
The physical picture of Ref.~\onlinecite{Wirnsberger2018} is that the various terms in our \eqnRef{equ:force-ideal} are all identified as forces (acting along the $z$ direction, which is the direction of $\nb$).
In particular, $T\Ac_T = \nb\cdot\nabla T$ and $T\Ac_\rho = (T/\rho)\nb\cdot\nabla\rho$ are referred to as `ideal forces'.
Moreover, the ideal force $T\Ac_T$ is identified as acting at the LJ centre, so it contributes to $f_\text{LJ}$ in \eqnRef{equ:wdfr-compare}, while the other ideal force $T\Ac_\rho$ is identified as acting at the centre of mass.

In more detail, the correspondence with Ref.~\onlinecite{Wirnsberger2018} is 
\beq
\begin{aligned}
f_{\rm LJ} &  \leftrightarrow  \langle \nb\cdot\bm{f} \rangle_{\bm r} - T\Ac_T,
\\
f_{\rm cm} &  \leftrightarrow  T\Ac_\rho,
\\
f_{\rm ext}& \leftrightarrow  F^\text{ext},
\end{aligned}
\label{equ:wdfr-map}
\eeq
where the forces of Ref.~\onlinecite{Wirnsberger2018} are on the left and the corresponding quantities in this work are on the right.  One sees that \eqnRef{equ:force-ideal} of this work corresponds to force balance $f_{\rm cm} + f_{\rm LJ}  + f_{\rm ext} = 0$, as in Ref.~\onlinecite{Wirnsberger2018}.  Combining \eqnRef{equ:wdfr-compare} with \eqnRef{equ:wdfr-map} and noting also $\langle\cos\theta(z)\rangle\leftrightarrow\langle\cos\theta\rangle_{\rb}$ recovers \eqnRef{equ:wdfr}.

We reiterate that the assignment of the ideal forces $T\Ac_T$ and $T\Ac_\rho$ to $f_{\rm LJ}$ and $f_{\rm cm}$ in \eqnRef{equ:wdfr-map} was justified in Ref.~\onlinecite{Wirnsberger2018} on phenomenological grounds.  A key result from this work is that the LEP distribution provides a microscopic basis for this choice.

Note also that all results in this work are for thermo-orientation in the absence of any external force $F^\text{ext}$.  These were called `NEMD runs' in Ref.~\onlinecite{Wirnsberger2018}, which also considered `$\nabla T$ runs', in which an external force was added in order that there was no density gradient $\Ac_\rho=0$, and `$\nabla\rho$ runs', where there was an external force and a density gradient but no temperature gradient.

The theory presented here can be applied in all cases.  However, for $\nabla\rho$ runs, the system is at equilibrium, and LEP predictions are exact (see Sec.~\ref{sec:lep}).  We have analysed $\nabla T$ runs, and have found that the absolute differences between LEP and EPB theory are very similar to NEMD runs (i.e.~those with no external force).  However, the total thermo-orientation effect is much larger if the external force is applied, so the deviations between LEP and EPB are less apparent in the $\nabla T$ runs.  The NEMD case is the one of primary interest, and it is also the most challenging for the theory, hence we focus on that case.

\section{Solutions to the one-particle Liouville equation}
\label{app:liouv}

This appendix shows that the LEP and EPB distributions are consistent with the Liouville equation \eqnRef{equ:liouville}.
Define an equilibrium-like distribution 
\beq
\PP_{\rm eq}(\eb,\,\pb,\,\Lb|\rb) = \frac{1}{T(\rb)^{d-1/2}}  \exp\mleft(\frac{-\KK}{T(\rb)}\mright) \frac{\PP_0}{Z_{\rm eq}},
\label{equ:PP-eq}
\eeq
where the kinetic energy $\KK$ is given by \eqnRef{equ:ke}.  We seek solutions to \eqnRef{equ:liouville} of the form $\PP = G \PP_{\rm eq}$ for some function $G$.  As usual, we work in the linear response regime; at zeroth order then $G=1$.
Dividing \eqnRef{equ:liouville} by $\PP_{\rm eq}$ and defining $G=\PP/\PP_{\rm eq}$ yields
\begin{align*}
0  = & \frac{1}{m} (\nb\cdot\pb)[\Ac_\rho   + \Ac_T (\KK - \KKb )/T ] G
\\
&  {} + I^{-1}  \boldnabla_\eb\cdot ((\Lb\times\eb) G)
 \\ 
& {} +  \boldnabla_\pb \cdot \left(\bfb G\right) - \frac{\pb \cdot (\nb F^\text{ext} +\bfb)}{mT} G
\\
& {} + \alpha\boldnabla_\Lb \cdot \left( (\eb\times\bfb) G \right) - \frac{\alpha}{IT} \Lb \cdot (\eb\times\bfb) G ,
\end{align*}
where $\KKb = (d-1/2) T$ is the mean kinetic energy.
We note that $\Ac_\rho$, $\Ac_T$, $\bfb$ ,$F^\text{ext}$, and $(G-1)$ are all small in the linear response regime where we work.
Keeping terms at first order and rearranging the order of some terms, we obtain
\begin{align}
0  = & \frac{1}{mT} (\nb\cdot\pb)[\Ac_\rho T  - F^\text{ext}   + \Ac_T (\KK - \KKb ) ]
 \nonumber \\ 
& {} +  \boldnabla_\pb \cdot \bfb + \alpha\boldnabla_\Lb \cdot(\eb\times\bfb)
 \nonumber \\ 
& {} + I^{-1} (\Lb\times\eb) \cdot \boldnabla_\eb G
 \nonumber \\ 
& {} - \frac{\pb \cdot\bfb}{mT}  - \frac{\alpha}{IT} (\Lb \times \eb)\cdot\bfb ,
\label{equ:liou-app}
\end{align}
where we have also rearranged some terms using vector product identifies, particularly $\boldnabla_\eb\cdot ((\Lb\times\eb) G) = (\Lb\times\eb) \cdot \boldnabla_\eb G$.

We note that the derivatives with respect to $\Lb$ and $\eb$ in \eqnRef{equ:liou-app} treat the orientation and angular momenta as vectors with three components, but we are restricting to cases where $\eb$ is a unit vector and $\eb\cdot\Lb=0$.
One may also verify this analysis with $\eb$ being represented explicitly as a point on the unit sphere, so that the vector $\boldnabla_\eb G$ is tangential to the sphere.
For the angular momentum, a consistent analysis requires that $\bfb$ is independent of $\Lb\cdot\eb$.

\subsection{Boltzmann solution (for $\Ac_T=0$)}

We first consider the case where there is no temperature gradient, but an external force $F^\text{ext}$ is applied.
The system is at equilibrium but the force does induce an alignment of the particles, as in the `$\nabla\rho$ runs' of Ref.~\onlinecite{Wirnsberger2018}.
It is useful to write
\beq 
G_{\rm Boltz} = 1 + \gamma_{\rm Boltz} (\nb\cdot\eb) 
\label{equ:GB}
\eeq
and 
\beq
 \bfb_{\rm Boltz} = f_{\rm Boltz} \nb 
 \label{equ:bfbB}
 \eeq
Here $\gamma_{\rm Boltz},f_{\rm Boltz}$ are constants, independent of $\eb$, $\pb$ and $\Lb$.  One may interpret $G$ as a first-order Taylor expansion of an orientational distribution $\PP_{\rm Boltz}(\nb) \propto \ee^{\gamma_{\rm Boltz} (\nb\cdot\eb)}$, where $\gamma_{\rm Boltz}$ is a free energy gain (normalised by the temperature) for alignment of $\eb$ with $\nb$.
Assuming (\ref{equ:GB},\ref{equ:bfbB}) and setting $\Ac_T=0$, consistency with \eqnRef{equ:liou-app} requires
$$
f_{\rm Boltz} = T \Ac_\rho - F^\text{ext}\qquad\text{and}\qquad \gamma_{\rm Boltz} = f_{\rm Boltz}.
$$
The form of $G_{\rm Boltz}$  also implies $\langle\cos\theta\rangle_\rb=\gamma/d$, so this is also consistent with Eqs~\eqref{equ:wdfr} and \eqref{equ:force-ideal}.

\subsection{LEP solution}
\label{app:lep-liouv}

In the Boltzmann case, \eqnRef{equ:liou-app} is linear in momenta.
In this case, $\bfb$ is independent of momenta.
However, if $\Ac_T\neq0$, then \eqnRef{equ:liou-app} contains terms cubic in momenta, due to the presence of the kinetic energy $\KK$.
In the LEP solution, one chooses $\bfb$ so that the term proportional to $\Ac_T\KK$ in \eqnRef{equ:liou-app} is cancelled by the term proportional to $\pb\cdot\bfb$.
This requires
\[
 \bfb_{\rm LEP} = \nb \left[ u_0 T + {\Ac_T} \left( \KK - \KKb \right)  \right] ,
\]
as in \eqnRef{equ:bfb-LEP}.
With this choice, the term in \eqnRef{equ:liou-app} proportional to $(\Lb \times \eb)\cdot\bfb$ generates an additional non-linear term that is cancelled by the term proportional to $\boldnabla_\eb G$, which requires
\beq 
G_{\rm LEP} = 1 +(\nb\cdot\eb) \left[\gamma + \alpha \Ac_T (\KK-\KKb)/T \right] .
\label{equ:G-LEP}
\eeq
This choice, together with \eqnRef{equ:bfb-LEP}, ensures the cancellation of all terms in \eqnRef{equ:liou-app} that are non-linear in momenta.
Next, one must choose $u_0$ and $\gamma$ such that the linear terms cancel and \eqnRef{equ:liou-app} is satisfied.
This requires that \eqnRef{equ:u-f} holds, and that $\gamma=\alpha(u_0-\Ac_T)$ is consistent with \eqnRef{equ:gamma-u-alpha}.
Finally, we note that \eqnRef{equ:G-LEP} is consistent with \eqnRef{equ:PP-LEP} only if $\alphat=\alpha$.
Hence, at leading order, Eqs~\eqref{equ:PP-LEP}, \eqref{equ:bfb-LEP}, \eqref{equ:u-f} and \eqref{equ:gamma-u-alpha} are together consistent with \eqnRef{equ:liou-app} and therefore with \eqnRef{equ:liouville}, as required.

\subsection{EPB solution}
\label{app:epb-liouv}

We now turn to the EPB distribution.  Recall that the angular velocity is $\dot\eb=I^{-1}(\Lb\times\eb)$.  From Eqs~\eqref{equ:pp-delta} and \eqref{equ:bfb-delta}, one has $G=G_{\rm LEP}+(\alpha \Ac_T)\updelta G$ and $\bfb=\bfb_{\rm LEP}+(\alpha \Ac_T)\updelta\bfb$.
Since  $G_{\rm LEP}$ and $\bfb_{\rm LEP}$ together already solve \eqnRef{equ:liou-app}, that equation may be rewritten as
\begin{align}
0  = %& \frac{1}{m} (\nb\cdot\pb)[\Ac_\rho   - F^\text{ext}   + \Ac_T (\KK - \KKb )/T ]
% \nonumber \\ 
&  \boldnabla_\pb \cdot \updelta\bfb + \alpha\boldnabla_\Lb \cdot(\eb\times\updelta\bfb)
 \nonumber \\ 
& {} + \dot\eb \cdot \boldnabla_\eb (\updelta G)
% \nonumber \\ 
%& 
 - \frac{1}{T} \left( \frac{\pb}{m} + \alpha\dot\eb \right) \cdot\updelta\bfb  .
\label{equ:liou-app-edot}
\end{align}
We write $\updelta G = \updelta G_\kappa + \updelta G_\lambda + \updelta G_\xi$ and similarly $\updelta\bfb=\updelta\bfb_\kappa + \updelta\bfb_\lambda$ (note $\updelta\bfb_\xi=0$).  The linearity of \eqnRef{equ:liou-app-edot} means that if each pair (such as $\updelta G_\kappa$, $\updelta\bfb_\kappa$) solves \eqnRef{equ:liou-app-edot} separately, then their sum is also a solution.

\subsubsection{$\kappat$ term}

We first consider 
$$
\updelta\bfb_\kappa = \kappat \left[ \eb(\nb\cdot\eb)  \frac{|\Lb|^2}{I\alpha} -  \pb \frac{ \nb\cdot \dot\eb  }{\chi }
-\frac{T}{\alpha} \left( \nb - \eb(\nb\cdot\eb) \right) \right]
$$
and 
$$
\updelta G_\kappa = - \kappat (\nb\cdot\eb) \left[ \frac{\pb}{\chi T}  \cdot \left(  \frac{\pb}{m} +\alpha\dot\eb \right)  +1 - \frac{d}{\chi} \right].
$$
In \eqnRef{equ:liou-app-edot}, we recall that derivatives with respect to $\eb$ are taken at fixed $\Lb$.  Hence, for any fixed vector $\bm{a}$, one has
\beq
\nabla_\eb^\mu (\dot\eb\cdot\bm{a}) = I^{-1} \nabla_\eb^\mu (\Lb\times\eb\cdot\bm{a})= I^{-1} (\bm{a}\times\Lb)^\mu .
\label{equ:diff-edot}
\eeq
Making use of this formula,
\begin{multline}
\dot\eb\cdot\boldnabla_\eb (\updelta G_\kappa) =   - \kappat (\nb\cdot\dot\eb) \left[\frac{\pb}{\chi T} \cdot \left(  \frac{\pb}{m} +\alpha\dot\eb \right)  +1 - \frac{d}{\chi} \right]
\\ - \alpha \kappat \frac{\nb\cdot\eb}{\chi I T} (\dot\eb\cdot\pb\times \Lb).
\label{equ:kappa-derivG}
\end{multline}
Note also that
\beq
\dot\eb\cdot\pb\times \Lb = I^{-1} ( \Lb\times\eb) \cdot (\pb\times\Lb) = -(\pb\cdot\eb)\frac{|\Lb|^2}{I}.
\eeq
Combining all these formulae and recalling $\chi=m\alpha^2/I$, one may verify that \eqnRef{equ:liou-app-edot} holds.

We observe that the functional forms of $\updelta\bfb$ and $\updelta G$ are constrained by the fact the second line of \eqnRef{equ:liou-app-edot} contains terms that are non-linear in momenta, which must all cancel.  In particular, this enforces that the non-linear terms in $\updelta G$ must be proportional to $\left(  \pb/m +\alpha\dot\eb \right)$, which is the same factor multiplying $\updelta\bfb$ in \eqnRef{equ:liou-app-edot}, and may be identified as the velocity $\dot\Rb_i$ of the LJ centre of the molecule.

\subsubsection{$\lambdat$ term}

The analysis of this case is similar to the previous one, except that the nonlinearity in $\updelta G$ includes terms quadratic in $\dot\eb$, instead of terms quadratic in $\pb$.  In particular
\beq
\updelta G_\lambda = - \lambdat (\nb\cdot\eb) \left[  \frac{m\alpha\dot\eb}{\chi T} \cdot \left(  \frac{\pb}{m} +\alpha\dot\eb \right)  - d \right],
\label{equ:dG-lambda}
\eeq
where one again notices the same factor of $\left(  \pb/m +\alpha\dot\eb \right)$.
Hence
\begin{multline}
\dot\eb\cdot\boldnabla_\eb (\updelta G_\lambda) =   - \lambdat (\nb\cdot\dot\eb) \left[  \frac{m\alpha\dot\eb}{\chi T} \cdot \left(  \frac{\pb}{m} +\alpha\dot\eb \right)  - d \right]
\\ - \alpha \lambdat \frac{\nb\cdot\eb}{\chi  IT} (\dot\eb\cdot\pb\times \Lb)  .
\label{equ:lambda-derivG}
\end{multline}
Also,
\beq
\updelta\bfb_\lambda = \lambdat \left[ \eb(\nb\cdot\eb)  \frac{|\Lb|^2}{I\alpha} -  m\alpha\dot\eb \frac{ \nb\cdot \dot\eb  }{\chi }   \right],
\label{equ:bfb-lambda}
\eeq
from which one has
$$
\eb\times \updelta\bfb_\lambda = -\lambdat m\alpha (\eb\times\dot\eb) (\nb\cdot\dot\eb)  .
$$
In order to evaluate the contribution $\boldnabla_\Lb\cdot(\eb\times \updelta\bfb_\lambda)$ in \eqnRef{equ:liou-app-edot}, it is important that $\bfb$ does not depend on the constant of motion $\eb\cdot\Lb$.
One uses
$(\eb\times\dot\eb) = I^{-1}[\Lb - \eb(\Lb\cdot\eb)]$ to write
$$
\eb\times \updelta\bfb_\lambda = -(\lambdat m\alpha/I^2) [\Lb - \eb(\Lb\cdot\eb)] (\nb\cdot\Lb\times\eb) .
$$
%This representation makes it clear that the force does not depend on the constant of motion $\Lb\cdot\eb$.  
The derivative of this expression can be evaluated as 
\beq
\alpha \boldnabla_\Lb \cdot ( \eb\times \updelta\bfb_\lambda )  = -\lambdat d (\nb\cdot\dot\eb)  .
\label{equ:div-L-lambda}
\eeq
Combining Eqs~\eqref{equ:dG-lambda}--\eqref{equ:div-L-lambda}, one may again verify \eqnRef{equ:liou-app-edot}.

\subsubsection{$\xi$ term}

The parameter $\xi$ does not appear in $\bfb$.
Instead one notes that
$$
\updelta G_\xi = \xi \frac{\pb\times\Lb\cdot\nb}{m\alpha T},
$$
for which $\boldnabla_\eb (\updelta G_\xi)=0$ and so \eqnRef{equ:liou-app-edot} holds trivially.

\section{Correlation functions for the LEP and EPB distributions}
\label{app:correl}

This Appendix shows how to compute expectation values with respect to the LEP and EPB distributions.
To first order one has
\begin{align*}
\PP_{\rm LEP} & = \PP_{\rm eq} [1 + G_{\rm LEP} ],
\\
\PP_{\rm EPB} & = \PP_{\rm eq} [1 + G_{\rm LEP} + \alpha\Ac_T \updelta G ],
\end{align*}
where $\PP_{\rm eq}$ was defined in \eqnRef{equ:PP-eq}, while $G_{\rm LEP}$ comes from \eqnRef{equ:G-LEP} and $\updelta G$ from \eqnRef{equ:delta-P}.
Hence (for any $A$) the EPB average is
$$
\langle A \rangle_\rb = \langle A [ 1 + G_{\rm LEP} + \alpha\Ac_T \updelta G ] \rangle_{\rm eq},
$$
where $\langle A \rangle_{\rm eq}$ indicates an average with respect to $\PP_{\rm eq}$.
Hence, all correlation functions can be reduced to averages with respect to this Gaussian distribution.   We summarise some general results before computing some specific correlations. The momentum $\pb$ is independent of all other variables, and
\begin{equation}\begin{split}
\langle p^\mu p^\nu \rangle_{\rm eq} & = mT\updelta^{\mu\nu},
\\
%\langle (\pb\cdot\nb)^2 |\pb|^2 \rangle_{\rm eq} & = (d+2)(mT)^2
%\nonumber\\
\langle |\pb|^4 \rangle_{\rm eq} - \langle |\pb|^2 \rangle^2_{\rm eq}& = 2d(mT)^2.
\end{split}\end{equation}
For the orientation one has
\begin{align}
\langle e^\mu e^\nu \rangle_{\rm eq} & = \updelta^{\mu\nu} / d,
\nonumber\\
\langle |\Lb|^2 \rangle_{\rm eq} & =  IT ( d-1),
\\
\langle |\Lb|^4 \rangle_{\rm eq} - \langle |\Lb|^2 \rangle^2_{\rm eq} & = 2(d-1)(IT)^2.\nonumber
\end{align}
The orientation and angular momentum are not completely independent (because $\Lb\cdot\eb=0$). One has
\begin{equation}\begin{split}
\langle e^\mu e^\nu |\Lb|^2 \rangle_{\rm eq} & = \langle e^\mu e^\nu \rangle_{\rm eq} \langle |\Lb|^2 \rangle_{\rm eq}, \\
% \langle |\Lb\times\eb|^2 \rangle_{\rm eq}   & = \langle  |\Lb|^2 \rangle_{\rm eq}
%\nonumber\\
\langle  (\Lb\times\eb)^\mu (\Lb\times\eb)^\nu \rangle_{\rm eq} & =  \frac{(d-1)IT}{d} \updelta^{\mu\nu}.
\end{split}\end{equation}
Since $\Lb$ and $\pb$ are independent, one also has
\begin{align}
\langle (\pb\times\Lb)^\mu (\pb\times\Lb)^\nu \rangle_{\rm eq} & = \frac{2 mIT^2 (d-1)}{d} \updelta^{\mu\nu}.
\label{equ:pLn-app}
\end{align}

\subsection{Average orientation and force}

Using the results above, one may now evaluate correlation functions. For example, the average orientation for the LEP distribution is
\begin{align}
\left\langle e^\mu \right\rangle_{\rb, \rm LEP}  & = \langle e^\mu (1 + G_{\rm LEP} ) \rangle_{\rm eq}
\nonumber \\ & = 
 \left\langle e^\mu (\nb\cdot\eb) [\gamma + \alpha \Ac_T (\KK-\KKb)/T] \right\rangle_{\rm eq}
\nonumber  \\ & = \frac{\gamma}{d} n^\mu .
 \label{equ:e-app}
\end{align}
To obtain the second line, we used the fact that $\PP_{\rm eq}$ is even in $\eb$, as well as \eqnRef{equ:G-LEP}.  To get to the third line, we note that $\KK$ is independent of the orientation.
To obtain the equivalent result for the full EPB distribution requires computation of the correction
$\alpha\Ac_T \langle e^\mu \updelta G \rangle_{\rm eq}$.  Performing the Gaussian integrals yields
\beq
\left\langle e^\mu \right\rangle_{\rb} =  \frac{\gamma+(\kappa-\lambda)\alpha\Ac_T}{d} n^\mu .
\label{equ:pol-app}
\eeq
Similarly one sees that (to leading order)
$$
\left\langle \bfb \right\rangle_{\rb} = \langle \bfb \rangle_{\rm eq} = u_0 T\nb .
$$
(The average $\langle\updelta\bfb\rangle_{\rm eq}$ of the EPB force  is zero.)  Using this formula with \eqnRef{equ:pol-app} and \eqnRef{equ:gamma-u-alpha} yields \eqnRef{equ:pol-epb}.

\subsection{Equipartition formulae}

Having explained the general method, we sketch the computations of other correlation functions.
To derive \eqnRef{equ:EEL}, which quantifies deviations from equipartition, one first considers the LEP case,
\begin{align*}
\left\langle e^\mu |\Lb|^2 \right\rangle_{\rb,\rm LEP}  & = \langle e^\mu |\Lb|^2 (1 + G_{\rm LEP} ) \rangle_{\rm eq}
\\ & = 
 \left\langle e^\mu |\Lb|^2 (\nb\cdot\eb) [\gamma + \alpha \Ac_T (\KK-\KKb)/T] \right\rangle_{\rm eq}
 \\ & = \frac{\gamma IT(d-1)}{d} n^\mu + \frac{\alpha \Ac_T IT(d-1)}{d} n^\mu.
\end{align*}
Using \eqnRef{equ:e-app} and rearranging, one has
\beq
\frac{1}{T(d-1)} \left\langle e^\mu \frac{|\Lb|^2}{I} \right\rangle_{\rb,\rm LEP}  - \langle e^\mu\rangle_{\rb,\rm LEP} = \frac{\alpha \Ac_T n^\mu}{d},
\label{equ:eel-lep}
\eeq
consistent with \eqnRef{equ:emu-L2-eq}.
The EPB correction to this quantity may be written as
\beq
\frac{\alpha\Ac_T}{T(d-1)} \left\langle e^\mu \left[ \frac{|\Lb|^2}{I} - T(d-1) \right] \updelta G \right\rangle_{\rm eq}.
\label{equ:dG-1}
\eeq
The only non-zero contribution to this average comes from terms proportional to $\updelta G_\lambda$.  The relevant term in $\updelta G_\lambda$ is $m\alpha^2 |\dot\eb|^2/(\chi T)=|\Lb|^2/(IT)$, where we used the definition of $\chi$. Noting that the other terms in $\updelta G_\lambda$ all have vanishing contributions to the average, \eqnRef{equ:dG-1} can be reduced to
$$
\frac{-\alpha\Ac_T}{T(d-1)} \left\langle e^\mu \left[ \frac{|\Lb|^2}{I} - T(d-1) \right] \lambdat (\nb\cdot\eb) \frac{|\Lb|^2}{IT} \right\rangle_{\rm eq}.
$$
This evaluates to
$
-2\lambdat n^\mu \alpha\Ac_T/d.
$
Adding this correction to the LEP result [\eqnRef{equ:eel-lep}] yields \eqnRef{equ:EEL}.

The other equipartition formula, \eqnRef{equ:EEP}, can be derived similarly.  The only term in $\updelta G$ that contributes to the EPB correction is $(\nb\cdot\eb)|\pb|^2/(m\chi T)$, which appears in $\updelta G_\kappa$.

\subsection{Other correlation functions}

To derive \eqnRef{equ:Qpar}, note that $\left\langle (f^\mu/d)  - e^\mu f^\nu e^\nu \right\rangle_{\rb, \rm LEP}=0$ and recall $\langle \updelta\bfb\rangle_{\rm eq}=0$.  Hence
\begin{align*}
\left\langle (f^\mu/d)  - e^\mu f^\nu e^\nu \right\rangle_{\rb} & = - \langle e^\mu (\updelta \bfb \cdot \eb) \rangle_{\rm eq}
\\
& = - \alpha\Ac_T \left\langle e^\mu (\nb\cdot\eb) ( \lambdat + \kappat ) \frac{|\Lb|^2}{I\alpha} \right\rangle_{\rm eq}
\\
& = -\Ac_T ( \lambdat + \kappat ) \frac{T(d-1)}{d} ,
\end{align*}
consistent with \eqnRef{equ:Qpar}.

To derive \eqnRef{equ:D-cor}, one writes
$$
\langle \pb\times\Lb\cdot\nb \rangle_{\rb} = \alpha\Ac_T \langle  (\pb\times\Lb\cdot\nb) \updelta G \rangle_{\rm eq}.
$$
There are several relevant contributions to $\updelta G$, all of which are linear in $\pb$ (otherwise the correlation function vanishes).
The contribution from $\updelta G_\xi$ is proportional to $\langle (\pb\times\Lb\cdot\nb )^2\rangle_{\rm eq}$, which is given in \eqnRef{equ:pLn-app}.  The contribution from $\updelta G_\kappa + \updelta G_\lambda$ is proportional to $\kappa+\lambda$ and also proportional to
$$
\langle (\nb\cdot\eb) (\pb\cdot\dot\eb) (\pb\times\Lb\cdot\nb )\rangle_{\rm eq} =
I^{-1} \langle (\nb\cdot\eb) (\pb\times\Lb\cdot\eb) (\pb\times\Lb\cdot\nb )\rangle_{\rm eq} ,
$$
which evaluates to $mT^2(d-1)/d$.  To show this, one may first average over the momentum $\pb$, which is independent of $\eb$ and $\Lb$, and then use vector product identities.

Following the same methodology, one may also show that
\begin{equation}\begin{split}
\frac{1}{T} \langle (\nb\cdot\eb)(\pb\cdot\dot\eb) \rangle_{\rb} & = -\Ac_T(\lambdat+\kappat-\xi)\frac{d-1}{d},\\
\frac{1}{T} \langle (\nb\cdot\dot\eb)(\pb\cdot\eb) \rangle_{\rb} & = -\Ac_T\xi\frac{d-1}{d}.
\label{equ:e-edot-np-app}
\end{split}\end{equation}
The second equation does not include $\kappa$ or $\lambda$ because
$$
\langle (\nb\cdot\dot\eb)(\nb\cdot\eb)(\pb\cdot\dot\eb)(\pb\cdot\eb) \rangle_{\rm eq}  = 0,
$$
which may be checked by performing the average over $\pb$ and then using $\eb\cdot\dot\eb=0$.
Writing $\Lb = I(\eb\times\dot\eb)$, one also sees that 
\beq
\pb\times\Lb\cdot\nb = I[ (\nb\cdot\eb)(\pb\cdot\dot\eb) - (\nb\cdot\dot\eb)(\pb\cdot\eb)] \;.
\label{equ:pxL-app}
\eeq   
Hence, taking the difference of the two lines of \eqnRef{equ:e-edot-np-app} provides an alternative derivation of \eqnRef{equ:D-cor}.
%and to consider the two contributions separately.

Finally, we note that writing \eqnRef{equ:dAdt} with $p_i^\mu$ replaced (on the left-hand side) by $(\bm{p}_i\cdot\eb_i)(\eb_i\cdot\nb) - (\bm{p}_i\cdot\nb)/d$ and using the equations of motion [\eqnRef{equ:eom}] yields
\beq
0 = \big\langle (\fb\cdot\eb)(\eb\cdot\nb) - (\fb\cdot\nb)/d + (\pb\cdot\dot\eb)(\eb\cdot\nb) + (\pb\cdot\eb)(\dot\eb\cdot\nb) \big\rangle_\rb,
\label{equ:epb-triv}
\eeq
where we dropped a term that vanishes in linear response.  Rearranging and using \eqnRef{equ:pxL-app}, we have
\beq
0 = \big\langle (\fb\cdot\eb)(\eb\cdot\nb) - (\fb\cdot\nb)/d + 2(\pb\cdot\dot\eb)(\eb\cdot\nb) - \frac1I (\pb\times\Lb\cdot\nb) \big\rangle_\rb.
\label{equ:epb-triv2}
\eeq
Since Eqs~\eqref{equ:epb-triv} and \eqref{equ:epb-triv2} follow directly from the equations of motion, both LEP and EPB ans\"atze must be consistent with them.  Note that \eqnRef{equ:epb-triv} relates the sum of the two correlation functions in \eqnRef{equ:e-edot-np-app} to the force-orientation correlation in \eqnRef{equ:Qpar} -- it is easily checked that the EPB ansatz is consistent with this constraint.
\end{appendix}

%merlin.mbs aipnum4-1.bst 2010-07-25 4.21a (PWD, AO, DPC) hacked
%Control: key (0)
%Control: author (8) initials jnrlst
%Control: editor formatted (1) identically to author
%Control: production of article title (0) allowed
%Control: page (0) single
%Control: year (1) truncated
%Control: production of eprint (0) enabled
%

%\bibliography{tmo_biblio}
\end{document}